%% file: NICE_arXiv_plain_June_2017.tex
\documentclass[12 pt]{article}

\usepackage{setspace,relsize}

\usepackage{moreverb}

\usepackage[margin=1.3in]{geometry}

\usepackage[pdftex]{lscape}
\usepackage{amscd}
\usepackage{amsmath,amsthm}
\usepackage{graphicx}
\usepackage{amsfonts}
\usepackage{amssymb}
\usepackage{setspace}
\usepackage{natbib}
\usepackage{mathrsfs}
\usepackage{multicol}
\usepackage{multirow}
\usepackage{pifont}
\usepackage{enumerate}
\usepackage{epic}
\usepackage{color}
\usepackage{booktabs}
\usepackage{authblk}
\usepackage{algorithm}
\usepackage{algpseudocode}
\usepackage{epstopdf}
\usepackage{subfig} 
\usepackage{newfloat} 
\usepackage{adjustbox} 
\usepackage{xcolor}
\DeclareMathOperator*{\argmin}{arg\,min}
\DeclareMathOperator*{\argmax}{arg\,max}

\input{def}

 \def\bfDelta{\mathbf \Delta}

\title{\large Network Induced Large Correlation Matrix Estimation}

\author{\small Shuo Chen$^1$\thanks{Correspondence to: shuochen@umd.edu}, Jian Kang$^2$, Yishi Xing$^1$, Yunpeng Zhao$^3$,   and Donald Milton$^4$ \\
\small $^1$ Department of Epidemiology and Biostatistics, University of Maryland, College Park, MD 20742, USA \\

\small $^2$  Department of Biostatistics, University of Michigan, Ann Arbor, MI 48109, USA
\\

\small $^3$  Department. of Statistics, George Mason University, Fairfax, VA 22030, USA 
\\
 
\small $^4$ Maryland Institute for Applied Environmental Health, University of Maryland, College Park, MD 20742, USA

}
\date{}

\begin{document}
\doublespacing \maketitle
 
\begin{abstract}
The correlation matrix of massive biomedical data (e.g. gene expression or neuroimaging data) often  exhibits a complex and organized, yet latent graph topological structure.  We propose a two step procedure that first detects the latent graph topology with parsimony from the sample correlation matrix and then regularizes the correlation matrix by leveraging the detected graph topological information. We show that the graph topological information guided thresholding can reduce false positive and false negative rates simultaneously because it allows edges to borrow strengths from each other precisely.  Several examples illustrate that the parsimoniously detected latent graph topological structures may reveal underlying biological networks and guide correlation matrix estimation. 
\end{abstract}

\emph{Keywords}: graph, large correlation matrix, network,  parsimony,  shrinkage, thresholding, topology.


\section{Introduction}

We consider a large data set $\mathbf X_{n \times p}$ with the sample size $n$ and the feature dimensionality of $p$.  The estimation of the covariance matrix $\boldsymbol \Sigma$ or correlation matrix $\mathbf{R}$  is fundamental to understand the inter-relationship between variables of the large data set $\mathbf X_{n \times p}$ (\citealp{Fan15}). 

 Various regularization methods have been developed to estimate the high-dimensional covariance matrix. For instance,  $\ell_1$ the penalized maximum likelihood  has been utilized to estimate the sparse precision matrix $\mathbf\Theta =\mathbf \Sigma^{-1}$
(\citealp{Friedman08}; \citealp{Ban08}; \citealp{Yuan07}; \citealp{Lam09}; \citealp{Yuan10}; \citealp{Cai11b}; \citealp{Shen12}). In addition, the covariance matrix thresholding methods have been developed to directly regularize the sample covariance matrix  (\citealp{Bickel08}; \citealp{Rothman09}; \citealp{Cai11a}; \citealp{Zhang10}; \citealp{Fan13}; \citealp{Liu14}). Similarly, the thresholding regularization techniques have also been applied to correlation matrix $\mathbf{R}$ estimation (\citealp{Sun06}; \citealp{Liu14}; \citealp{Cui16}).
 \citet{MH} and  \citet{Witten11} point out that the two sets of methods are naturally linked regarding vertex-partition of the whole graph and estimate of the graph edge skeleton.

Graph notations and definitions are often used to describe the relationship between the $p$ variables of $\mathbf X_{n \times p}$ (\citealp{Yuan07}; \citealp{MH}).  A finite undirected graph $G=\{V,E\}$ consists two sets, where the vertex set $V$ represents the variables $\textbf{X}=(X_1, \cdots, X_p)$ with $|V|=p$ and the edge set $E$ denotes relationships between the vertices.  Let $e_{i,j}$ be the edge between nodes $i$ and $j$. 
Then $e_{i,j} $ is an connected edge if nodes $i$ and $j$ are genuinely  correlated in $G$. 
Under the sparsity assumption, the regularization algorithms assign most edges as unconnected, and $G$ may be decomposed to a set of maximal connected subgraphs 
(\citealp{Witten11}; \citealp{MH}). 

Many recent works estimate the covariance matrix by taking the graph topological structure into account.  For example, \citet{Witten11}, \citet{Hsieh12}, \citet{Witten15} utilize the diagonal block structure and  \citet{Bien16} use the \textit{banding} structure to improve the estimation of the covariance matrix. In many biomedical high-dimensional data sets, we find interactions between biological features (e.g. genes or neural processing units) often exhibit an interesting organized network  graph topological pattern which consists a number of block/community subgraphs and a large random subgraph (see Figure 1).  In this paper we develop a two step procedure that 1) first detects the graph topological structure \textit{parsimoniously} and 2) then estimates the correlation matrix by leveraging the information of this graph topological structure.


Motivated by multiple practical biomedical large data analyses, we consider a graph topological structure as $G=G^1 \cup G^0$ where the subgraph $G^1=\cup_{c=1}^{C_1} G_c$ is a stochastic block model structure and $G^0=\cup_{c=1}^{C_0} G^0_c$ ($G^0_c$ is a singleton only consisting one node) can be considered as an Erd\"{o}s-R\'{e}nyi random graph (i.e. $G$ is a mixture model and we refer it as the $G^1 \cup G^0$ mixture model). Thus, the $G^1 \cup G^0$ mixture model is a special case of the stochastic block model, which contains many singletons and  a number of communities (\citealp{Bickel09}; \citealp{Karrer11}; \citealp{Zhao11}; \citealp{Choi12}; \citealp{Newman12}; \citealp{Lei14}).  However, the data sets are often noisy and thus the conventional clustering algorithms can not easily identify such structures (\citealp{Witten15}). Therefore, we propose a new parsimonious algorithm to effectively estimate the $G^1 \cup G^0$ mixture structure, which is robust to false positive noises (edges). Our new approach impose an penalty term on the size of edges  all $\{ G_c \}$ such that include most  highly correlated  edges in communities of $G^1$ in communities with the minimum sizes of edges. That the new penalty term reduce the impact noises on the community estimation is due to the fact that the (sample) false positive highly correlated edges are often distributed in a random pattern rather in a community structure.

In step two, we estimate the large  correlation   by using the detected $G^1 \cup G^0$ mixture model graph topological structure. Specifically, we perform  thresholding for edges within and outside communities adaptively by using Bayes factors, and thus detected graph topology serves as prior knowledge. In this way, the decision of thresholding  an edge is made upon considering both this edge's magnitude and the its \textit{neighborhood}  via the detected graph topological information. Therefore, our network based thresholding strategy allows edges to borrow strength from each other while avoiding the the computationally  difficult step of estimation of the \textit{covariance of edges} (i.e. correlation of correlations). Different from methods of \citet{Hsieh12} and \citet{Witten15} which mainly focus on the edges within block components, we utilize information of edges from both inside and outside diagonal blocks. We name the whole graph topology information guided regularization strategy \textbf{N}etwork \textbf{I}nduced \textbf{C}orrelation matrix \textbf{E}stimation (NICE).  

The NICE method makes three contributions: i) we propose a new penalized objective function that is well-suited to estimate latent graph topological structures and very robust to false positive noises and we develop efficient algorithms to solve the objective function;  ii) we fuse the graph topological information and threholding decision making procedure to simultaneously reduce false positive and false negative discovery rates; iii) new statistical theories are developed.   In addition, the detected topological structures not only assist to estimate the dependency relationship between each pair of variables, but also can reveal interesting underlying biological networks.

 The paper is organized as follows. Section 2 describes the NICE algorithm, followed by  theoretical results in Section 3. In Sections 4, we perform the simulation studies and model evaluation/comparisons and  we apply our method to a mass spectrometry proteomics data set. Concluding remarks are summarized in Section 5.

\section{Methods}
We consider the sample covariance $\textbf{S}$ and  sample correlation matrix $\widehat{\textbf{R}}$=diag$(\textbf{S})^{-1/2}$ $\textbf{S}$ diag$(\textbf{S})^{-1/2}$ as our input data (\citealp{Sun06}; \citealp{Liu14}; \citealp{Fan15}). We may directly perform hard thresholding on the sample correlation matrix to estimate \textbf{R} by using $  R^{\mathcal{T}}_{i,j} = \{ \widehat R_{i,j}I(|\widehat R_{i,j}|>T) \}$ without understanding the underlying network structure, where $T$ is a pre-specified or calculated threshold. 
However, applying the universal regularization  rule (even when optimal $T$ is provided) to each element (or column) may introduce numerous false positives and false negatives due to various noises from the sample data.  Therefore, we propose to leverage the information from the latent topological structure of the correlation matrix (i.e. graph $G$)  to assist the decision making process. 
\begin{figure}[!tbp]
  \centering
  \subfloat[The truth: two networks]{\includegraphics[width=0.5\textwidth]{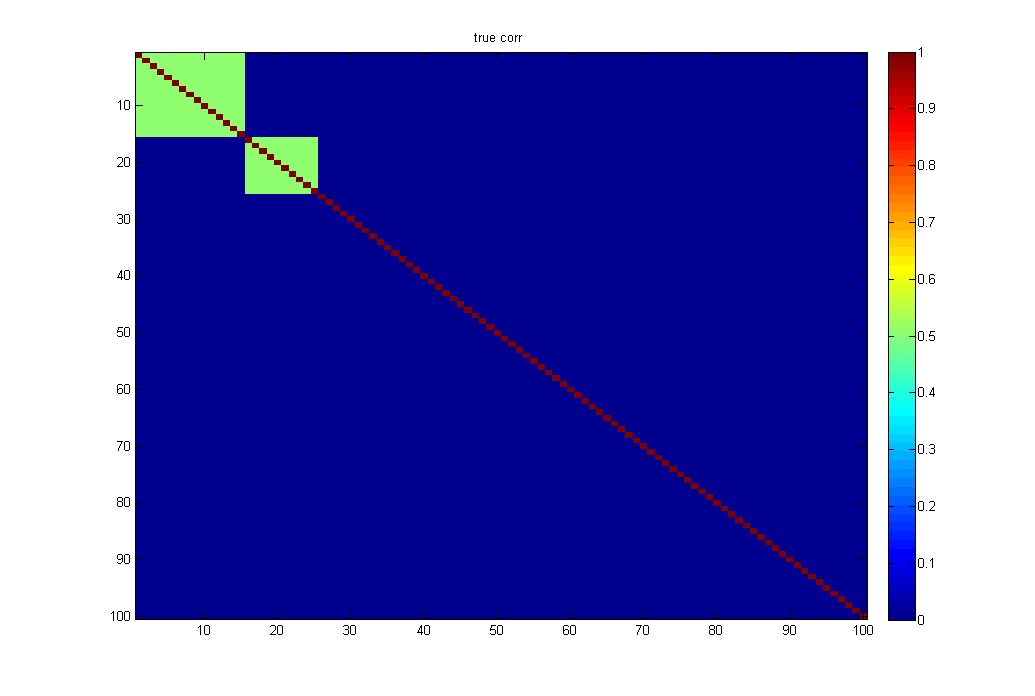}\label{fig1:f1}}
  \hfill
  \subfloat[Shuffling the order of nodes]{\includegraphics[width=0.5\textwidth]{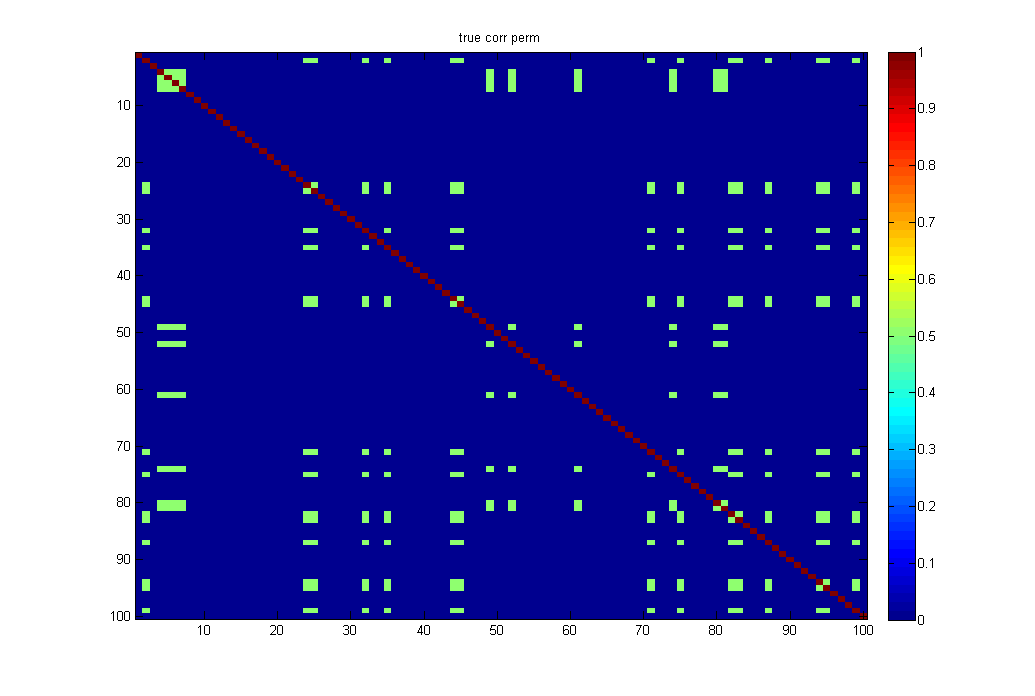}\label{fig1:f2}}
  \hfill
  \subfloat[The input data for NICE]{\includegraphics[width=0.5\textwidth]{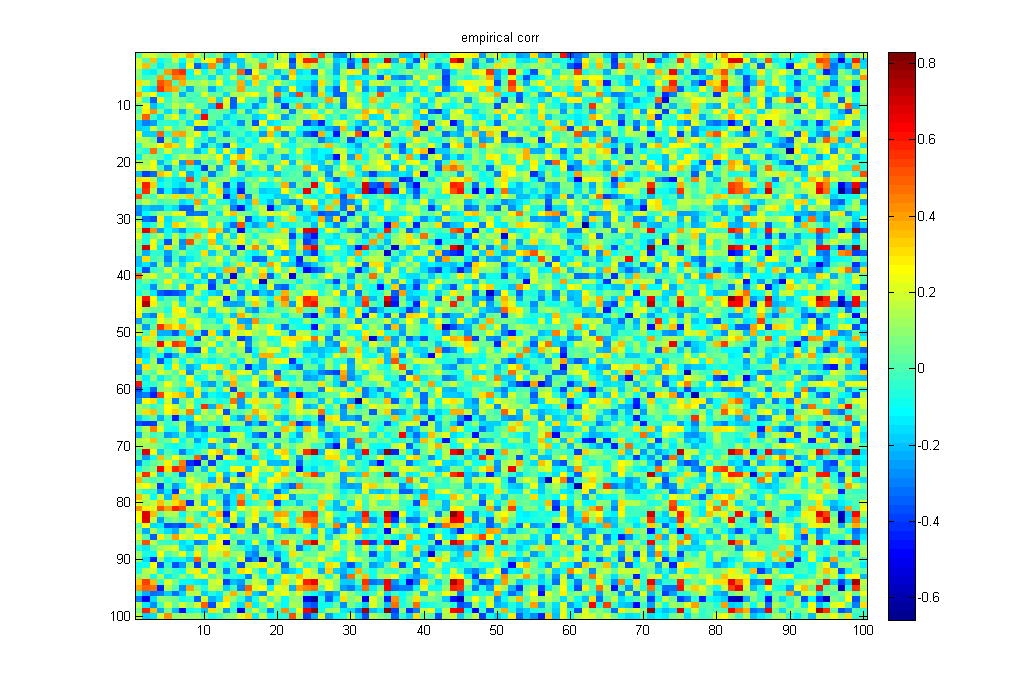}\label{fig1:f3}}
  \hfill
  \subfloat[Network detection results]{\includegraphics[width=0.5\textwidth]{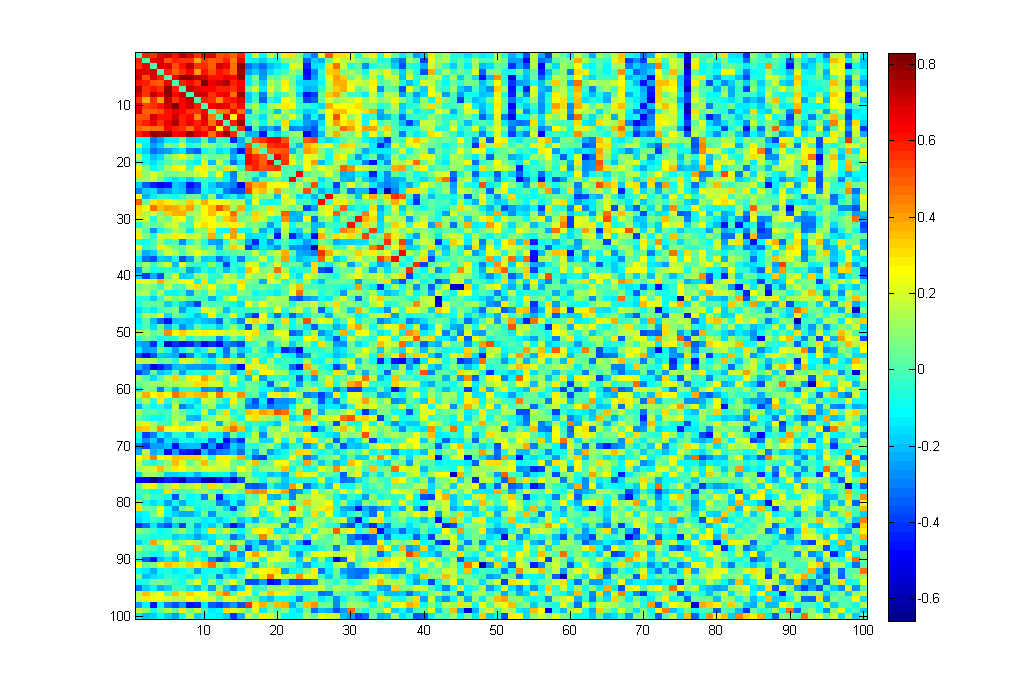}\label{fig1:f4}} 
  \caption{An example of a network induced covariance matrix: $|V|$=100 nodes and $|E|$=4950 edges, there are two networks (a) and in practice they are implicit (b) especially hard to recognize when looking at the sample covariance matrix (c); however, with the knowledge/estimation of topological network structures detected by NICE (d) the regularization strategy should take them into account.}
\end{figure}

The NICE method consists two steps: i) 
we first detect the latent topological structure of $G=G^1 \cup G^0 $ mixture in $G$ by applying the rule of parsimony  
ii)we then apply empirical Bayes based thresholding to the sample correlation matrix guided by the detected graph topology.  


\subsection{Parsimonious estimation of latent networks from sample correlation matrix}

We first define the weight matrix  \textbf{W}  based on the empirically estimated correlation matrix $\widehat{\textbf{R}}$. An entry $w_{i,j}$ of \textbf{W} can be a transformed correlation coefficient between variables $i$ and $j$  that corresponds to the edge $e_{i,j}$ in $G$, for example, Fisher's Z transformation.  $w_{i,j}$ is often a continuous metric. In Appendix, we describe an empirical Bayes based procedure to calculate $w_{i,j}$ as a metric between 0 and 1. \textbf{W} is only used for the latent network detection rather than the regularization step.

We assume that  $G$ includes induced complete subgraphs (community networks) as shown in Figure \ref{fig1:f1}, the edges within the networks are more likely to be connected than edges outside networks. However, in practice this topological structure is latent. The sample correlation matrix  has no explicit graph topological structure (\ref{fig1:f3}).  By optimizing the objective function \ref{fm1} we can recognize the latent graph topological structures (\ref{fig1:f4}). Next, we perform permutation tests to evaluate the statistical significance of each $G_c$, and the statistically significant subgraphs $\{G_c\}$ will be used to assist the estimation of the correlation matrix in the following step. 

We aim to identify the latent   $G^1 \cup G^0$ mixture   structure from \textbf{W} by using  penalized optimization. The heuristic is to identify a set of subgraphs $U= \cup_{c=1}^C G_c $ that maximizes the sum of weights of edges in the union set with minimum subgraph sizes. The penalty term is used to avoid mis-recognizing the network structure due to false positive noises.   Formally, we propose an objective function   
\begin{align}
\underset{C,  \{G_c\} }{\argmax} \sum_{c=1}^C \exp\{ \log(\sum (w_{i,j}|e_{i,j} \in G_c))-\lambda_0 \log(|E_c|) \},
\label{fm1}
\end{align}
with following definitions and conditions:
 
1.  $G_c$ ($c =1, \cdots, C$) is a clique subgraph that $G_c=\{V_c, E_c\}$ and $|V_c|\geq 1$;
  
2.  the size of the a subgraph $G_c$ is determined by the number of edges $|G_c|=|E_c|$;

3.  $\cup_{c=1}^C V_c=V$, $\cap_{c=1}^C V_c=\emptyset$ and $\cup_{c=1}^C E_c \subseteq E$.

The objective function is non-convex and difficult to be directly solved. We develop iterative algorithm to optimize $C$ and  $\{G_c\}$. In Appendix, we provide the detailed derivation and optimization algorithms (including choosing the tuning parameter $\lambda_0$), which links \ref{fm1} to a spectral clustering related objective function (\citealp{von07}; \citealp{Newman12}). The number of subgrphs $C$ is considered to be related to the penalty because   $C=1$ leads to $\cup_{c=1}^C E_c  =  E $ and $C=|V|$ indicates $\cup_{c=1}^C E_c  = \emptyset$. Because of the penalty term, the objective function often selects a relatively large $\hat C$ value and include many $G_c$ as singletons by graph size shrinkage. The objective function is  well-suited to capture topological structure from sample correlation matrix while being less affected by the false positive noises by implementing the new penalty term.
  
Therefore, implementing optimization to \ref{fm1} yields estimates of the network topological structure underlying within the large correlation matrix, which can be used to guide the decision making procedure of correlation matrix thrsholding.

\subsection{Graph topology oriented correlation matrix threshohlding}

To estimate the correlation matrix $\textbf{R}$, we perform graph topology guided  thresholding on the sample correlation matrix $\widehat{\textbf{R}}$ by using Bayes factors. Let $z_{i,j}$ be the Fisher's Z transformed sample correlation coefficient of $\widehat R_{i,j}$ and it follows a mixture distribution that $z_{i,j} \sim \pi_0 f_0(z_{i,j})+\pi_1 f_1(z_{i,j})$. 

\textit{Universal thresholding}

Without considering prior information of the topology structure, the universal thresholding  can be applied(\citealp{Bickel08}). For instance, an empirical Bayes framework implements a Bayes factor based via the  (\citealp{Efron04}, \citealp{Shafer05}). The hard-thresholding rule  is often employed for this purpose (\citealp{Cai11a}, \citealp{Fan15}), which sets an edge to zero unless   
\begin{align*}
\frac{\mathrm{P}(\delta_{i,j}=1|z_{i,j})}{\mathrm{P}(\delta_{i,j}=0|z_{i,j})} = \frac{f_1(z_{i,j})\pi_1}{f_0(z_{i,j})\pi_0}\geqslant T,
\end{align*}
 $T$ is a constant that is linked to local $fdr$ cutoff, and $\pi_0$ and $\pi_1$ are the proportions of null and non-null distributions correspondingly.  For example, $T=4$  is equivalent to the cutoff of local $fdr$ of 0.2 (\citealp{Efron07}). For instance, given $\pi_0=0.9$ and  $\pi_1=0.1$, the universal decision rule is that an edge is thresholded when BF=$\frac{f_1(z_{i,j})}{f_0(z_{i,j})} \leq  36$. In practice,  $\pi_0$ and $\pi_1$ are estimated based on the distribution of the statistics (e.g. $z_{i,j}$) and the Bayes factor cut-off is updated accordingly. 

It has been well documented that the Bayes factor inferential models could adjust the multiplicity by adjusting the prior structure (\citealp{Jeffreys61}; \citealp{Kass95}; \citealp{Scott06}; \citealp{Efron07}; \citealp{Scott10}). The prior odds are tuned to control false positive rates, and a larger $\pi_0$ ($\pi_0  \rightarrow 1$) or a distribution of $\pi_0$ with larger mean leads to more stringent adjustment that may cause both low false positive discovery rates and high false negative discovery rates.  \citealp{Scott06} suggest a prior distribution with median value around 0.9 and \citealp{Efron07} estimates $\pi_0$ by using an empirical Bayes model. 

However, there has been a long term challenge for all universal regularization methods (e.g. shrinkage or thresholding): the trade-off between false positive and false negative findings. Moreover, edges may be dependent on each other in an organized topological pattern and the mass univariate edge inference (universal regularization) ignoring the dependency structure may not estimate the large covariance and correlation matrix effectively and efficiently. Yet, the direct estimation of the dependency structure between edges is challenging and sometimes not feasible. We propose one possible solution by leveraging latent graph topology to guide thresholding and account for the \textit{dependency} between edges. The detected topological structure can seamlessly fuse into the empirical Bayes thresholding framework  as prior knowledge and provides precise neighborhood information that allows edges to borrow strengths for each other. 

\textit{Network based thresholding}

In a network induced correlation matrix, an edge with sample correlation value $z_{i,j}$ is more likely  to be truly connected within than outside a network community because the within community `neighbor' edges are more connected. Thus, we incorporate the topological location information of an edge into the regularization procedure. We first calculate the prior odds (of being truly connected) for edges within and outside  community networks separately by: 

\begin{align*}
\begin{split}
& \theta_{in} =\frac{\mathrm{P} (\delta_{i,j}=0|e_{i,j} \in  G_c, \forall c)}{\mathrm{P} (\delta_{i,j}=1|e_{i,j} \in  G_c,\forall c)}=\frac{\pi_0^{in}}{\pi_1^{in}} , \\
 & \theta_{out}=\frac{\mathrm{P} (\delta_{i,j}=0|e_{i,j} \notin  G_c, \forall c)}{\mathrm{P} (\delta_{i,j}=1|e_{i,j} \notin  G_c,\forall c)}=\frac{\pi_0^{out}}{\pi_1^{out}},   
\end{split}
\end{align*}

Clearly,  the within  community edges are more connected by  and thus $\pi_1^{in} >\pi_1> \pi_1^{out} $  and $\pi_0^{out} >\pi_0> \pi_0^{in} $, and $\theta_{out} \geq \theta_{all} \geq \theta_{in}$.

Let edges inside and outside of the detected communities follow different distributions: 
\begin{align*}
\begin{split}
 &z_{i,j}|e_{i,j}   \in  G_c \sim \pi_0^{in} f_0(z_{i,j})+\pi_1^{in} f_1(z_{i,j}); \\
 &z_{i,j}|e_{i,j} \not\in    G_c \sim \pi_0^{out} f_0(z_{i,j})+\pi_1^{out} f_1(z_{i,j}).
 \end{split}
\end{align*}

The proportions are different for inside and outside network, and overall edges, yet we assume that the null  $f_0(z_{i,j})$  and non-null $f_1(z_{i,j})$ distributions are identical.
By using the identified the latent community networks where edges are more correlated in step one, we propose the network based  thresholding rule:


If $e_{i,j}   \in  G_c$,
\[ \widehat{R}_{i,j}^{\mathcal{T}} = \left\{ \begin{array}{ll}
            \widehat R_{i,j}  & \mbox{if  $BF_{i,j} = \frac{\hat{f}_1(z_{i,j})}{\hat{f}_0(z_{i,j})} \geq T\cdot\hat{\theta}_{in}$};\\
        0 & \mbox{otherwise}.

        \end{array} \right. \]

else if $e_{i,j} \not\in    G_c$,
 
\[ \widehat{R}_{i,j}^{\mathcal{T}} = \left\{ \begin{array}{ll}
            \widehat R_{i,j}  & \mbox{if $BF_{i,j} =\frac{\hat{f}_1(z_{i,j})}{\hat{f}_0(z_{i,j})} \geq T\cdot\hat{\theta}_{out}$};\\
        0 & \mbox{otherwise}.

        \end{array} \right. \]
Equivalently, the we provide estimate of the edge  set  $\widehat{E}$ by using: 
\begin{align}
\begin{split}
  \widehat \delta_{i,j}^{0,in}&= I\left(  \frac{\hat{f}_1(z_{i,j})}{\hat{f}_0(z_{i,j})} \geq T\cdot\hat{\theta}_{in}\right) \\
   \widehat \delta_{i,j}^{0,out}&= I\left(  \frac{\hat{f}_1(z_{i,j})}{\hat{f}_0(z_{i,j})} \geq T\cdot\hat{\theta}_{out}\right),\\
\label{fm8}
\end{split}
\end{align}

where $\delta^0_{i,j}$ is an indicator variable that $\delta^0_{i,j}=1$ when variables $i$ and $j$ are correlated with each other, otherwise $\delta^0_{i,j}=0$.

The detected graph topology provides the prior knowledge of the `neighborhood' and `location' of an edge, which accordingly allow graph topology guided thresholding while accounting for dependency between edges. 
A community network is analogous to a neighborhood (spatial closeness) of \textit{edges} with explicit boundaries and  edges within the neighborhood could borrow power from each other. Many statistical models are developed based on this idea, for example, the Ising prior and conditional autoregressive (CAR) model (\citealp{Besag95}). Nevertheless, unlike data in spatial or imaging statistics the sample correlation/covariance matrix of large biomedical data sets often include no available information about the exact spatial location or closeness. Thus, our detected graph topological structure provides a new pathway of regularization/statistical inferences on `edges' by accounting for the dependency structure  based on (detected) latent graph topological `closeness'. 


 We empirically estimate $\widehat{\theta}_{in}$ and $\widehat{\theta}_{out}$   from the data. {First, let all edges that are Fisher's Z transformed sample correlation coefficients in $\widehat{\textbf{R}}$ follow a mixture distribution $f(z_{i,j})=\pi_0^{all}f_0(z_{i,j})+\pi_1^{all}f_1(z_{i,j})$. We estimate {$\widehat{\pi}_0^{all}, \widehat{\pi}_1^{all},\widehat f_0, \widehat f_1$} similarly to local $fdr$ by algorithms used in \citep{Efron04}. Next, we estimate $\widehat{\pi}_0^{in}$ for in-network edges $e_{i,j} \in G_c$. Since $\widehat f_0, \widehat f_1$ are estimated in the previous step, the only parameter to estimate in $f^{in}(z_{i,j})=\pi_0^{in}f_0(z_{i,j})+\pi_1^{in}f_1(z_{i,j})$ is  $\widehat{\pi}_0^{in}= 1- \widehat{\pi}_1^{in}$. We simply implement the maximum likelihood estimation and then obtain $\widehat{\theta}_{in}=\widehat{\pi}_0^{in}/\widehat{\pi}_1^{in}$.  
For edges outside of networks ($z_{i,j}$ that $e_{i,j} \notin G_c$ ) $f^{out}(z_{i,j})=\pi_0^{out}f_0(z_{i,j})+\pi_1^{out}f_1(z_{i,j})$  we estimate $\widehat{\pi}_0^{out}$    and calculate $\widehat{\theta}_{out}=\widehat{\pi}_0^{out}/\widehat{\pi}_1^{out}$ by following the same procedure. 
In general, our graph topological structure detection algorithm produces a very small odds ratio $\widehat \theta^{in}/ \widehat \theta^{out}$  when the informative edges are distributed in an organized pattern. Thus, the choice of  $T$ has a small impact on the decision making process.
Uncovering  graph topological structure is important to understanding the interactive relationships between multivariate variables (nodes) and the dependency between edges.  We show that the detected topological structure can also provide prior knowledge to assist large covariance/correlation matrix regularization and estimation.  The network based regularization approach utilizes the additional yet latent graph structure information and reduces false positive and negative discovery rates simultaneously (shown in section 4).   We summarize the NICE algorithm of both steps in Algorithm \ref{NICEalg} in the Appendix.

\def\mF{\mathrm{Fis}}

\section{Theoretical Results}
We start with some notations for the theoretical development. Let $X_{i,k}$ be the observed data on node $i$ for subject $k$, for $i = 1, \ldots, p_n$ and $k = 1,\ldots, n$ with mean zero and unit standard deviation.  Recall that $\bfR = \{ R_{i,j} \}$ is the true correlation matrix of interests with $\mathrm{Cor}(X_{i,k},X_{j,k}) = R_{i,j}$ and $\widehat\bfR = \{ \widehat R_{i,j}\}$ be the correlation matrix estimator.  Let $\mF(x) = \log\left\{(1+x)/(1-x)\right\}/2 $ be the Fisher's Z transformation.  Let $z_{i,j} = \mF(\widehat R_{i,j})$ and $\mu_{i,j} = \mF(R_{i,j})$, and $R_{i,j} = 0$ if and only if $\mu_{i,j} = 0$. Let $f_{i,j}(\cdot)$ be the density function of $z_{i,j}$. Let $\phi(x) = \exp(-x^2/2)/\sqrt{2\pi}$ be the standard normal probability density function. Let $\delta^0_{i,j} = I[R_{i,j} \neq 0]  = I[\mu_{i,j}\neq 0] = I[e_{i,j} \in G]$ indicate whether or not $e_{i,j}\in G$.  Let $q_n = \sum_{1\leq i < j\leq p_n} \delta^0_{i,j}$, 
\begin{eqnarray*}
f_{0}(z) = \frac{\sum_{i < j} (1-\delta^0_{i,j}) f_{i,j}(z) }{\sum_{i<j}(1-\delta^0_{i,j})},\qquad \mbox{ and } \qquad f_{1}(z) = \frac{\sum_{i<j} \delta^0_{i,j} f_{i, j}(z) }{\sum_{i<j}\delta^0_{i,j}}. 
\end{eqnarray*}
Let $f(z)$ denote the actual distribution of $\{ z_{i,j}\}_{1\leq i < j\leq p_n}$. 
\begin{eqnarray*}
f(z) = \pi_0 f_0(z) + \pi_1 f_1(z),
\end{eqnarray*}
where 
$$\pi_0 = \frac{p_n(p_n - 1) - q_n }{p_n(p_n - 1)}\quad \mbox{and} \quad \pi_1 = 1 - \pi_0.$$

Given data $\{z_{i,j}\}_{i<j}$, suppose $\widehat f_c(\cdot)$  be an estimator for $f_c(\cdot)$ for $k= 0, 1$, and $\widehat \pi_0$ is an estimator for $\pi_0$ with $\widehat \pi_1 = 1-\widehat \pi_0$. For any $T>0$ and $\widehat\bfR$,  define the $\mathrm{NICE}$ thresholding operator   $\mathrm{NICE}(\widehat\bfR; T) = \{\mathrm{NICE}(\widehat R_{i,j}; T)\}_{i<j}$. Specifically, 
$$
\mathrm{NICE}(\widehat R_{i,j};T)= \left\{\begin{array}{cc}
\widehat R_{i,j}, & \frac{\widehat f_1(z_{i,j})}{\widehat f_0(z_{i,j})} > \frac{\widehat\pi_0}{\widehat\pi_1}T,\\
0 &\frac{\widehat f_1(z_{i,j})}{\widehat f_0(z_{i,j})} \leq \frac{\widehat\pi_0}{\widehat\pi_1}T.
\end{array}\right.
$$

\subsection{Conditions}
The following conditions are needed to facilitate the technical details, although they may not be the weakest conditions. 
\begin{condition}\label{con:data}
We consider the following conditions on the data $\bfX = (X_{i,k})$.
\begin{enumerate}
\item Data are centered around zero with unit variance. i.e $\mE[X_{i,k}] = 0$ and $\mVar[X_{i,k}] = 1$.
\item Data are uniformly bounded.  That is, there exists a constant $M>0$ such that 
$$\mathrm{P}[|X_{i,k}| < M] = 1,$$
\item The Pearson's correlation estimator is computed by 
$$\widehat R_{i,j} = \frac{1}{n}\sum_{c=1}^n X_{i,k} X_{j,k},$$
\item The population level correlation satisfy
$$|R_{i,j}| = |\mE[X_{i,k} X_{j,k}] |< 1.$$
\end{enumerate}
for all $1 \leq i, j \leq p_n$ and $k = 1,\ldots, n$. 
\end{condition}
\begin{condition}\label{con:mu}
There exist  constants $c_0>0$ and $\tau > 0$, such that 
$$ \mu_{\inf} = \inf_{i < j}\{|\mu_{i,j}| : \delta^0_{i,j} = 1\}  = c_0 n^{-1/2+\tau}.$$
\end{condition}

\begin{condition}\label{con:p_q}
Let $\tau>0$ be the same constant in Condition \ref{con:mu}. Then 
$$\log\left(p_n\right) = o(n^{2\tau}).$$
\end{condition}

\begin{condition}\label{con:pi}
Suppose there exists $0.5 < \pi_0 < 1$, such that
$$\lim_{n\to\infty} \pi_0 = \pi_0 \mbox{ and } \lim_{n\to\infty} \pi_1 = 1-\pi_0.$$
\end{condition}

\begin{condition}\label{con:est}
Given data $\{z_{i,j}\}_{i<j}$, suppose $\widehat f_c(\cdot)$  be a consistent estimate for $f_c(\cdot)$ for $k= 0, 1$. Suppose $\widehat \pi_0$ is consistent estimates for $\pi_0$. Specifically, for any $\epsilon > 0$,
\begin{eqnarray*}
\lim_{p_n \to \infty} \mathrm{P}\{\|\widehat f_c(\cdot)- f_c(\cdot)\|_2 + |\widehat\pi_0 - \pi_0|>\epsilon\} = 0.
\end{eqnarray*}
where $\|\cdot\|_2$ be the $L_2$ norm for the function, which is defined as $\|f\|_2 = \int_{\mbR} \{f(x)\}^2 \md x$. 

\end{condition}

\begin{condition}\label{con:fp}
Let  $\omega=(\sum_{c=1}^C |V_c| \times (|V_c|-1)/2 )/ (|V| \times (|V|-1)/2) $  the proportion of edges inside community networks and $\int_{z_0}^ \infty f(z_{i,j})= F(z_0)$. $z_0$ is the universal threshold cut-off value , $z_{0,in}$ is the within networks threshold cut-off value, $z_{0,out}$ is the within networks threshold cut-off value.
$$\frac{ F_0(z_0) - F_0(z_{0,out}) }{ F_0(z_{0,in}) - F_0(z_0) } > \frac{\omega  \pi^{in}_0}{(1-\omega)  \pi^{out}_0}$$
$$\frac{ F_1(z_0) - F_1(z_{0,out}) }{ F_1(z_{0,in}) - F_1(z_0) } < \frac{\omega  \pi^{in}_1}{(1-\omega)  \pi^{out}_1}.$$

This condition is generally valid for network induced correlation matrix because by implementing the parsimonious estimation of network topological structure $f^{in}$ is distinct from $f^{out}$. Thus, we have $\pi^{in}_0\ll\pi^{out}_0$ and $\pi^{in}_1\gg\pi^{out}_1$, and condition holds.  
\end{condition}

\subsection{Tail Probability Bounds}
By Berry-Esseen theorem and Taylor expansion,  it is straightforward to show the following lemma: 
\begin{lemma}\label{lem:distr_fun}
For any $i < j$ , 
\begin{eqnarray*}
\lim_{n\to\infty} \sup_{z\in \mbR}\left| f_{i,j}(z) - \sqrt{n}\phi\{\sqrt{n}(z-\mu_{i,j})\}\right| = 0.
\end{eqnarray*}
\end{lemma}
In addition, we also need to study the probability bound of the $z_{i,j}$. Specifically, we have the following lemma:
\begin{lemma}\label{lem:prob_bound}
Suppose Condition \ref{con:data} holds. For all $1\leq i,j\leq p_n$, there exists a constant $K>0$ and $N>0$,  for and $n>N$, and any $\epsilon > 0$, we have
$$\mathrm{P}[\sqrt{n}|z_{i,j}-\mE[z_{i,j}]| > \epsilon] \leq \exp\left(- K\epsilon^2\right).$$
\end{lemma}

By Lemma \ref{lem:distr_fun} and Condition \ref{con:data}, we can uniformly approximate $f_1(z)/{f_0(z)}$, which is stated in the following lemma:

\begin{lemma}\label{lem:f_ratio}

\begin{eqnarray*}
\lim_{n\to\infty} \sup_{z\in\mbR}\left|f_{0}(z) - \sqrt{n}\phi\{\sqrt{n} z\} \right| = 0,\\
\lim_{n\to\infty} \sup_{z\in\mbR}\left|f_{1}(z) - \frac{1}{q_n}\sum_{i<j} \delta^0_{i,j}\sqrt{n}\phi\{\sqrt{n}(z-\mu_{i,j})\}\right| = 0,
\end{eqnarray*}
and 
\begin{eqnarray*}
\lim_{n\to\infty}\sup_{z\in\mbR}\left|\frac{f_1(z)}{f_0(z)} - \frac{\pi_0}{\pi_1} \frac{\sum_{i<j}\delta^0_{i,j}\phi\{\sqrt{n}(z-\mu_{i,j})\}}{(p_n(p_n-1)/2 - q_n)\phi(\sqrt{n} z)}\right| = 0,
\end{eqnarray*}
\end{lemma}

\begin{lemma}\label{lem:norm_bound}
Suppose Conditions \ref{con:mu}--\ref{con:pi} hold. There exist constants $C_0>0$ and $C_1>0$ such that for any $i< j$ and any $T>(1-\pi_0)/\pi_0$, there exists $N_T>0$, for all $n>N_T$, we have
when $e_{i,j} \in G$, then 
$$\mathrm{P}\left[\frac{\sum_{i'<j'}\delta^0_{i',j'}\phi\{\sqrt{n}(z_{i,j}-\mu_{i',j'})\}}{\{p_n(p_n-1)/2 - q_n\}\phi(\sqrt{n}z_{i,j})}\leq T\right] \leq  \exp(-C_1 n^{2\tau} ).$$
And when $e_{i,j} \notin G$, then 
$$\mathrm{P}\left[\frac{\sum_{i',j'}\delta^0_{i',j'}\phi\{\sqrt{n}(z_{i,j}-\mu_{i',j'})\}}{\{p_n(p_n-1)/2 - q_n\}\phi(\sqrt{n}z_{i,j})}> T\right] \leq C_2\exp(-C_0 n^{2\tau}).$$
Note that $N_T$ depends on $T$ but it does not depend on $i$ and $j$. 
\end{lemma}

\subsection{Selection and Estimation Consistency}
We construct the population level selection indicator $\widetilde\delta_{i,j}(T)$ and the selection indictor estimator $\widehat \delta_{i,j}(T)$ and discuss their properties in Lemmas \ref{lem:e_bound} and \ref{lem:e_hat_bound} respectively. 
\begin{lemma}\label{lem:e_bound}
Suppose Conditions \ref{con:mu}--\ref{con:pi} hold. For all $i < j$ and any $T>(1-\pi_0)/\pi_0$, let $$\widetilde\delta_{i,j}(T) = I\left[{\pi_1f_1(z_{i,j}) \over \pi_0 f_0(z_{i,j})} > T\right].$$
 Then there exist $N_T>0$, $C_3>0$ and $C_4>0$ such that for any  $n > N_T$, 
\begin{eqnarray*}
\mathrm{P}\{\widetilde\delta_{i,j}(T) \neq \delta^0_{i,j}\} \leq C_3 \exp(-C_4 n^{2\tau}).
\end{eqnarray*}
where $N_T$ depends on $T$ but not on $i$ and $j$, and $\tau>0$ is the same constant in Condition \ref{con:mu}. 
\end{lemma}


\begin{lemma}\label{lem:e_hat_bound}
For any $T>(1-\pi_0)/\pi_0$ and any $i<j $, let
 $$\widehat\delta_{i,j}(T) = I\left[{\widehat \pi_1\widehat f_1(z_{i,j}) \over \widehat \pi_0\widehat f_0(z_{i,j})}> T \right].$$ 
 Then there exists $N_T>0$ such that for all $n>N_T$, 
 $$\mathrm{P}[\widehat \delta_{i,j}(T)\neq \delta^0_{i,j}] \leq C_3 \exp(-C_4 n^{2\tau}), $$
 where the constants $C_3$, $C_4$ and $\tau$ are the same as the ones in Lemma \ref{lem:e_bound}. 
\end{lemma}

We establish the selection consistency and estimation consistency in the following two theorems respectively. 
\begin{theorem}(Selection Consistency) \label{thm:selection_consistency}
Suppose Conditions \ref{con:data} -- \ref{con:est} hold. Denote by $\bfDelta_0 = \{\delta^0_{i,j}\}_{i<j}$ all the edge indicators. 
 For any $T>(1-\pi_0)/\pi_0$, let $\widehat{\bfDelta}(T) = \{\widehat \delta_{i,j}(T)\}_{i<j}$, then there exists $N_T>0$ for all $n > N_T$, 
$$\mathrm{P}\{\widehat{\bfDelta}(T)  = \bfDelta_0\} \geq 1 - \frac{C_3}{2}p_n(p_n-1)\exp( - C_4 n^{2\tau}).$$
where the constants $C_3$, $C_4$ and $\tau$ are the same as the ones in Lemma \ref{lem:e_bound}. Furthermore, 
$$\lim_{n\to\infty}\mathrm{P}\{\widehat{\bfDelta}(T)   = \bfDelta_0\}  = 1.$$
\end{theorem}

\begin{theorem}(Estimation Consistency) \label{thm:estimation_consistency}
Suppose Conditions \ref{con:data} -- \ref{con:est} hold and the constant $\tau$ in Condition \ref{con:est} satisfies  $0<\tau<1/2$.  For any $\epsilon > 0$ and any $T> (1-\pi_0)/\pi_0$, we have
$$\lim_{n\to\infty}\mathrm{P}[\|\mathrm{NICE}(\widehat\bfR; T) - \bfR\|_\infty> \epsilon] = 0,$$
where $\|\bfM\|_{\infty}$ is the $L^{\infty}$ norm, i.e. $\|\bfM\|_{\infty}  = \max_{1\leq i, j\leq p_n} |m_{i,j}|$ for any matrix $\bfM = (m_{i,j})$. 
\end{theorem}

\subsection{Reduced false positive and negative discovery rates by using NICE thresholding}

\begin{theorem}\label{thm:FPN}
Suppose Condition  \ref{con:fp}  hold, we have both 1) $E(\sum_{i<j} I(\widehat \delta_{ij}^{NICE}=1 | \delta_{ij}=0)) \leq E(\sum_{i<j} I(\widehat \delta_{ij}^{Univ}=1 | \delta_{ij}=0))$
  the expected false positively thresholded edges by using the graph topology oriented thresholding  (NICE) method   are less than the universal thresholding method; 2) $E(\sum_{i<j} I(\widehat \delta_{ij}^{NICE}=0 | \delta_{ij}=1)) \leq E(\sum_{i<j} I(\widehat \delta_{ij}^{Univ}=0 | \delta_{ij}=1))$, the expected false negatively thresholded edges by using the graph topology oriented thresholding (NICE) method are less than the universal thresholding method. 
  
\end{theorem}
%
%
%
%

\section{Data Analysis}
\subsection{Data examples} \label{data}
We  apply the NICE method to two publicly available high-dimensional biomedical data sets. By using these two examples, we show that the latent $G^1 \cup G^0$ mixture structure widely exists in data across platforms (e.g. proteomics, genomics, and imaging data, yet due to space limitation we only demonstrate two data types).  

\subsubsection{Proteomics data}

The first example is matrix-assisted laser desorption ionization time of flight mass spectrometry (MALDI-TOF MS) proteomics data from human 288 subjects (\citealp{MALDI07}). The data assess the relative abundance of  peptides/proteins in human serum. Each raw mass spectrum consists roughly 70,000 data points. After preprocessing steps including registration, wavelets denoising, alignment, peak detection, quantification, and normalization (\citealp{Chen09}), 184 features are extracted to represent the most abundant protein and peptide features in the serum. Each feature is located at a distinct m/z value that could be linked to a specific peptide or protein with some ion charges (feature id label). The original paper  utilizes the proteomics data to enhance  understanding of lung cancer pathology at the molecular level. In this paper, we focus on estimating the correlation matrix to investigate interactive relationships between these features.

We apply   NICE  to detect  correlated peptide/protein networks and estimate correlation matrix based on  the Fisher's Z transformed  sample correlation matrix (Figure \ref{fig2:f1}).   
First, the penalized objective function \ref{fm1} is implemented to capture the latent $G^1 \cup G^0$ mixture structure. The estimation results are $\hat{C}=77$, and that seven significant community networks ($G_1$) are detected  and the rest are singletons ($G_0$) (see Figure \ref{fig2:f2}). Figure \ref{fig2:f2}   reorders features Figure \ref{fig2:f1} by the detected topological structure. Generally, features within networks are more correlated than features outside networks

 We show that the distributions of edges inside and outside networks in Figure \ref{fig2:f5}. Clearly, $f^{in}$ and $f^{out}$ show distinct distributions, and $f^{out}$ is close to the null distribution for which all edges are not connected. We estimate $\hat{\pi}_0^{all}=0.78$, $\hat{\pi}_0^{out}=0.83$, and $\hat{\pi}_0^{in}=0.001$.
Then, we apply the network based thresholding  to estimate $\hat{E}$ and the  correlation matrix. The estimate $\hat{E}$ and thresholding rule $\{ \hat{\delta}_{i,j} \}$ are shown in Figure \ref{fig2:f6}.   The network detection results  provide informative inferences of the interactive relationship between these proteomics features.  In this data example, each network represents a group of related protein and peptides that can be confirmed by proteomics mass spectrometry literature. For example, the most correlated network three consists a list of proteins of normal and variant hemoglobins with one and two charges (\citealp{Lee11}) including normal hemoglobins $\alpha$ and $\beta$ with one charge and two charges (at m/z 15127, 15868, 7564, and 7934). The highly correlated networks of biomedical features may   provide guidance to identify a set of biomarkers for future research that allow to borrow power between each other. 

\begin{figure}[!htp]
  \centering
  \subfloat[Sample correlation]{\includegraphics[width=0.5\textwidth]{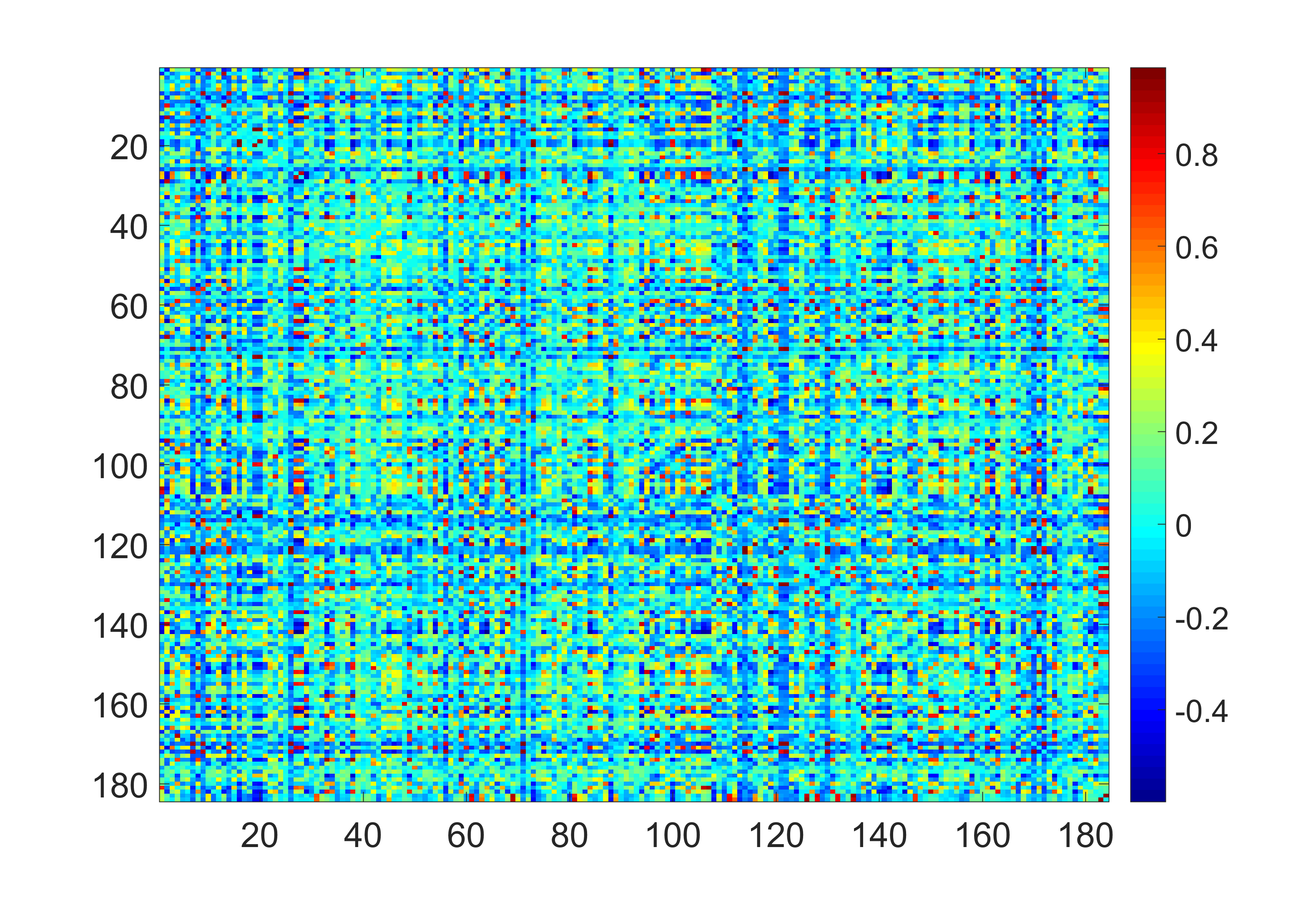}\label{fig2:f1}}
  \hfill
  \subfloat[Detecting latent $G^1 \cup G^0$ mixture structure ]{\includegraphics[width=0.5\textwidth]{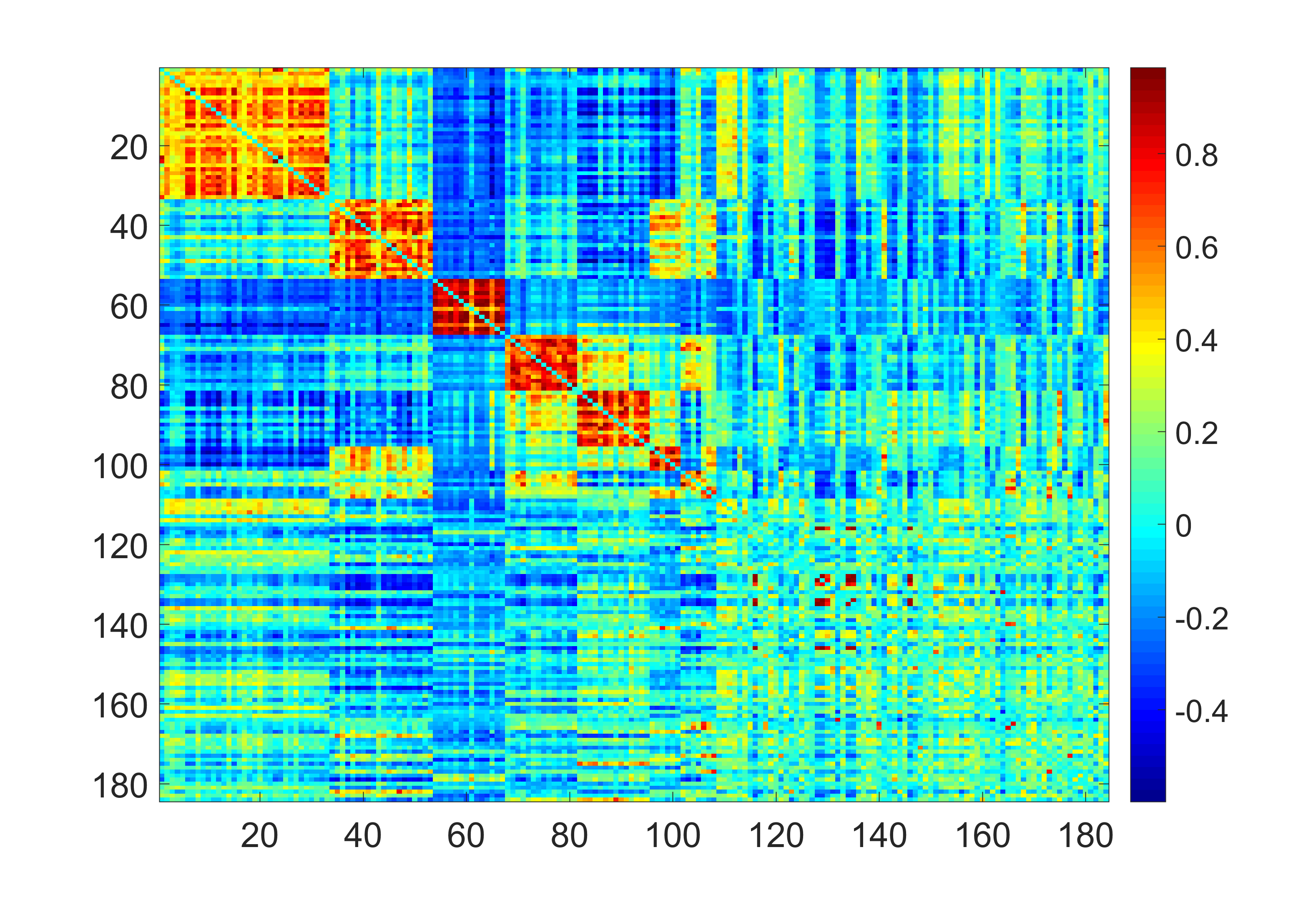}\label{fig2:f2}}
  \hfill
  \subfloat[Edges inside and outside networks]{\includegraphics[width=0.5\textwidth]{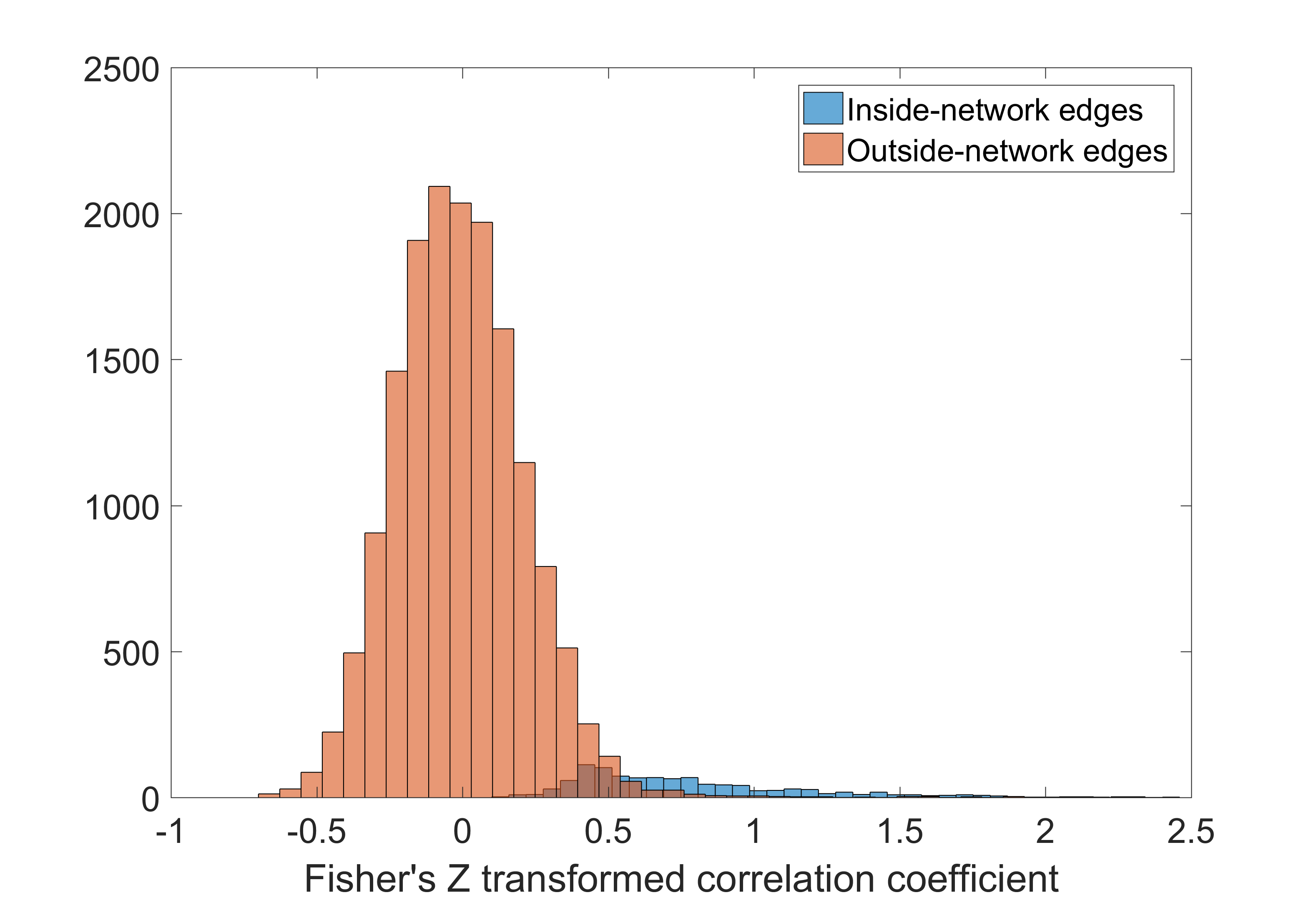}\label{fig2:f5}} 
  \hfill
  \subfloat[Estimated edge set $\hat{\textbf{E}}$]{\includegraphics[width=0.5\textwidth]{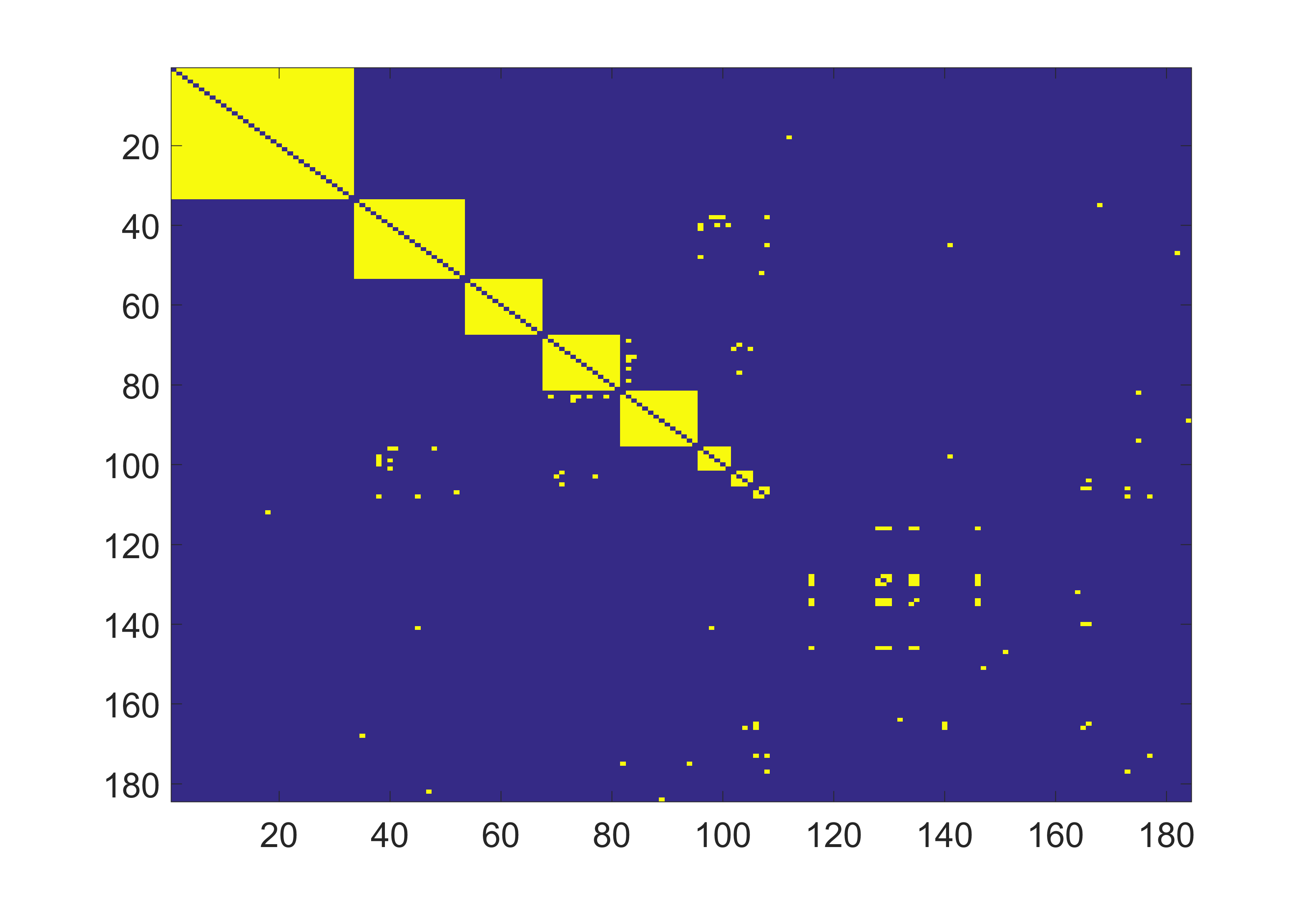}\label{fig2:f6}} 
  \caption{Application of the NICE to the example data set one. (a) is the heatmap of sample correlation matrix; (b) demonstrates the latent  $G^1 \cup G^0$ mixture structure by reordering the variables in the heatmap; (c) shows the distributions of edges inside and outside the networks; (d) is the estimated $\hat{E}$ based on the NICE thresholding.}
\end{figure}

\begin{figure}
\centering 
\caption{\textit{Glasso} results for Example Data 1: it shows that \textit{Glasso}  may false negatively regularize edges to zero in networks (with the sparsity assumption).}
 
  \subfloat[Correlation heatmap in the order of detected communities]{\includegraphics[width=0.55\textwidth]{Corr_vicc77.png}} \\

  \subfloat[\textit{glasso} results ]{\includegraphics[width=0.55\textwidth]{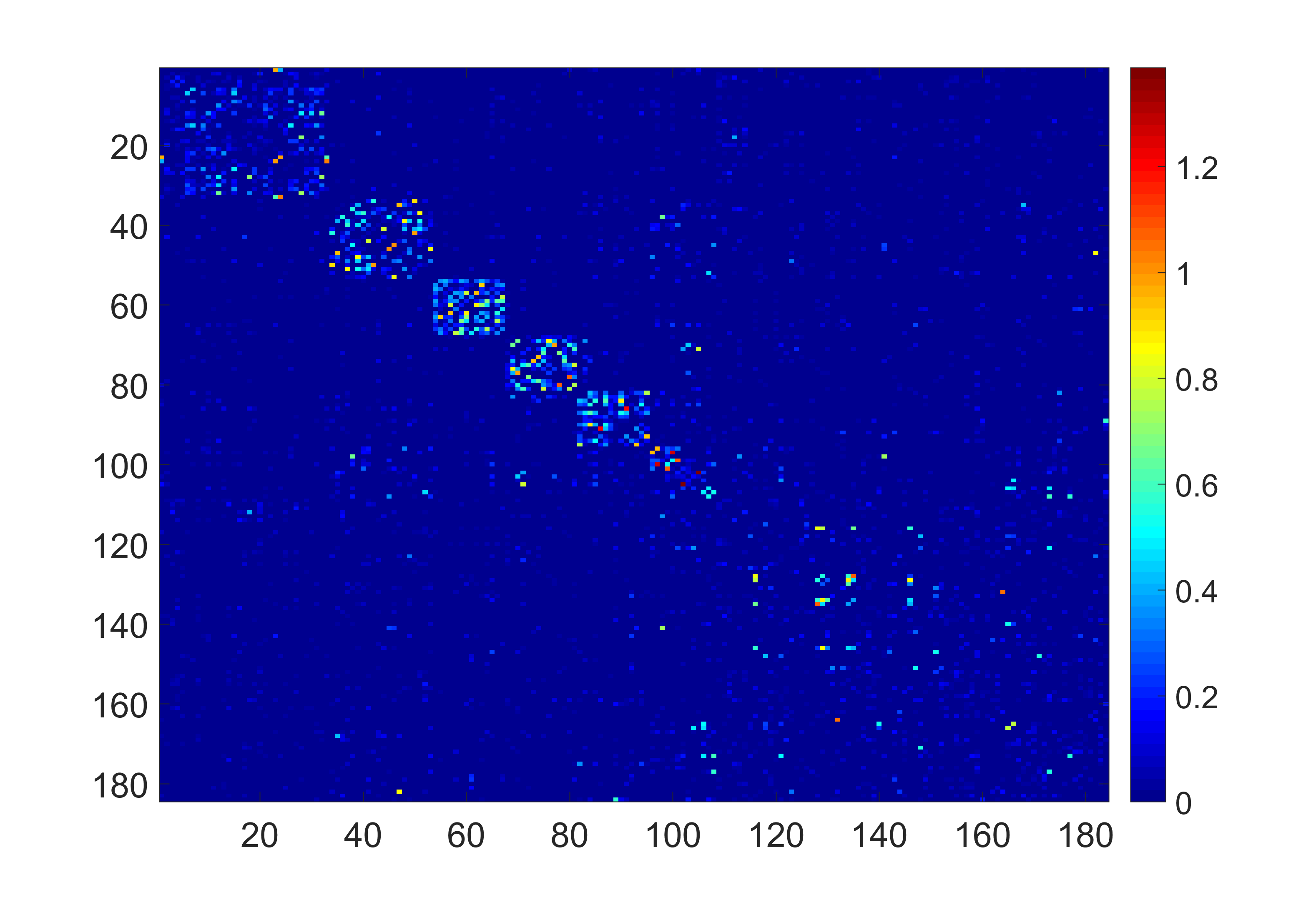}\label{fig3:f2}}
   \hfill

\end{figure}

\subsubsection{Gene expression data}

The second data example is gene expression profiling data based on Affy Human Genome U133A 2.0 array.  The data is publicly available at Gene Expression Omnibus (GEO) with accession code: GSE17156, GSE30550, GSE52428 and used by Dream Challenge (see more details at \url{https://www.synapse.org/#!Synapse:syn5647810/wiki/399110}).  Blood samples were collected for 110 healthy controls at baseline.  We focus on 1924 gene expression features that are commonly observed in human blood, and normalized data is used for analysis. The input data for our model is a $1924 \times 1924$ sample correlation matrix (Figure \ref{fig3:f1}). The sample correlation matrix show no explicit organized topological structures. By applying  the penalized objective function in \ref{fm1}, we identify the latent  $G^1 \cup G^0$ mixture structure (Figure \ref{fig3:f2}). Note that Figure \ref{fig3:f2} is a isomorphic graph to Figure \ref{fig3:f1} with reordered nodes. With $\hat{C}=613$, four large networks and a long list of singletons and small networks (with 2 or 3 nodes) are detected because of the penalty term. 

Figure \ref{fig3:f3} shows that edges inside and outside community networks follow distinct distributions. We estimate $\hat{\pi}_0^{all}=0.84$, $\hat{\pi}_0^{out}=0.99$, and $\hat{\pi}_0^{in}=0.05$. The distribution of edges outside of community networks is also close to the null distribution of non-connected edges, whereas the distribution of edges inside networks again centers around 0.5. By applying the network guided thresholding, we obtain  the estimated  correlation matrix and $\hat{E}$ as shown in figure \ref{fig3:f4}.

Interestingly, the latent $G^1 \cup G^0$ mixture structure shows in both data examples, which can also be identified in large data from many other platforms including neuroimaging activation and connectivity data, DNA methylation data, and etc. (\citealp{Chen16a}). In addition, for most of these data sets the inside and outside network edge distributions tend to be distinct with $f^{out}$ close to the null distribution and $f^{in}$ centers around 0.5. This further verifies that our assumptions of mixture distribution and condition \ref{con1} are generally valid.  

In comparison, when we apply   existing methods (e.g. \textit{glasso}), the latent $G^1 \cup G^0$ mixture structure can not identified based on the estimated covariance or inverse covariance matrix. Many inside network edges are (false negatively) regularized to zero (see Supplementary Materials).

\begin{figure}[!htp]
  \centering
  \subfloat[Sample correlation]{\includegraphics[width=0.5\textwidth]{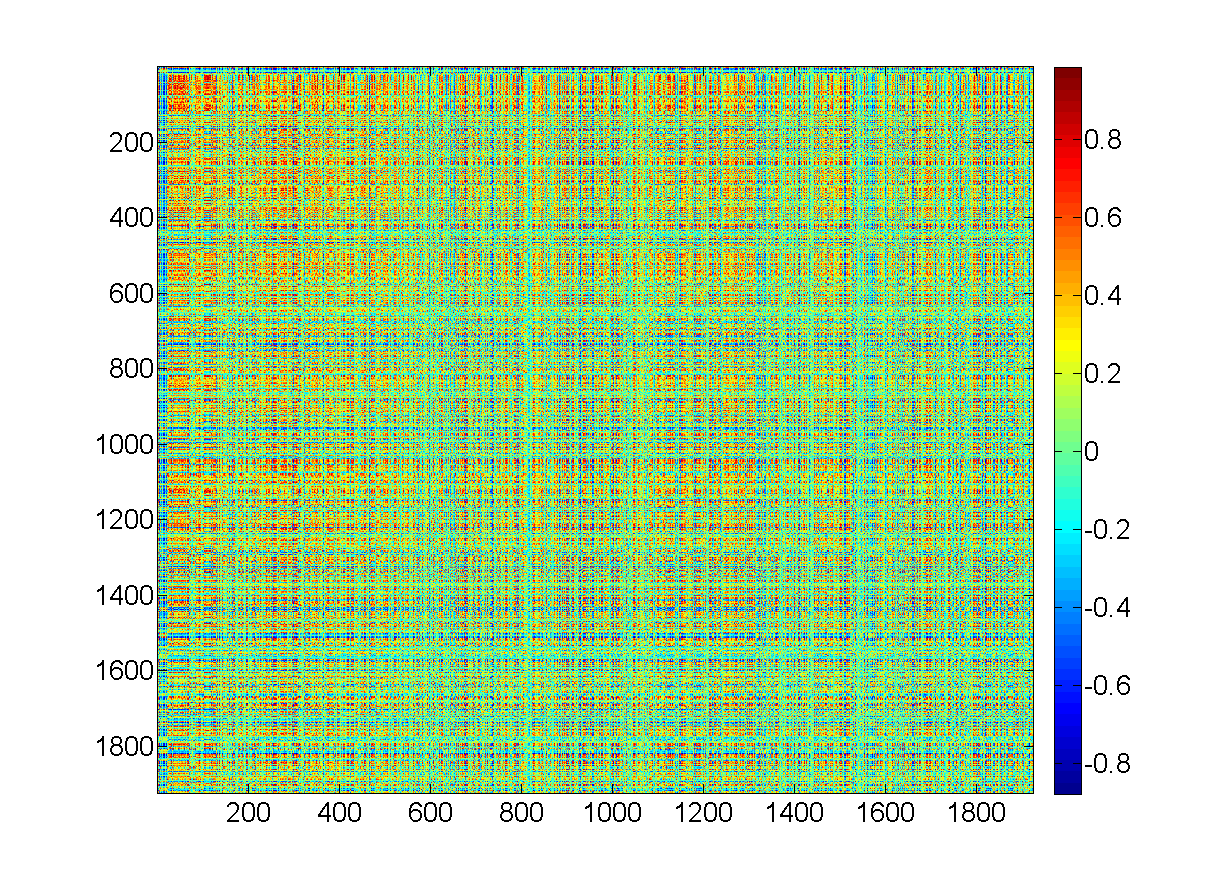}\label{fig3:f1}}
  \hfill
  \subfloat[Weight matrix \textbf{W} ]{\includegraphics[width=0.5\textwidth]{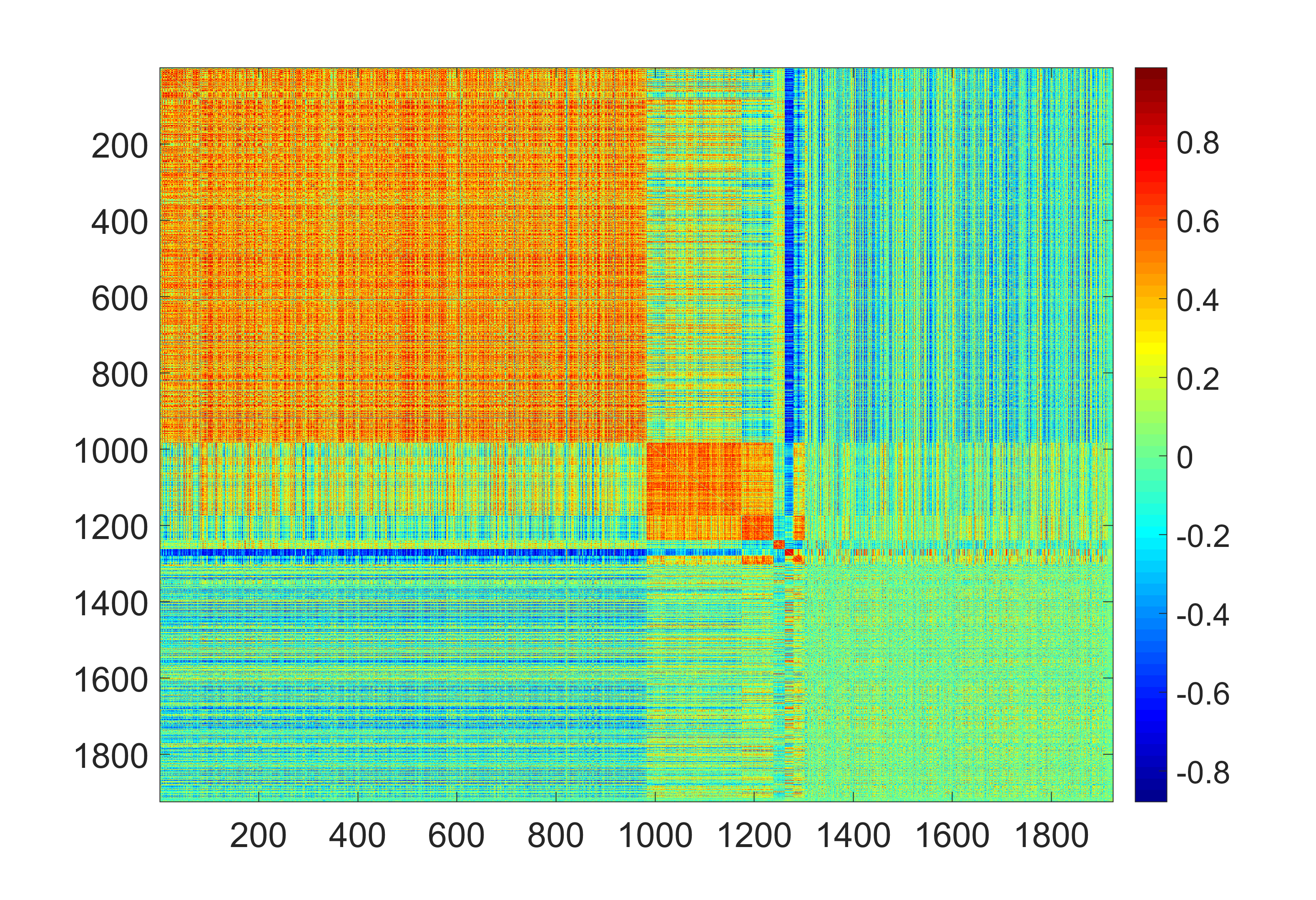}\label{fig3:f2}}
  \hfill

  \subfloat[Edges inside and outside networks]{\includegraphics[width=0.5\textwidth]{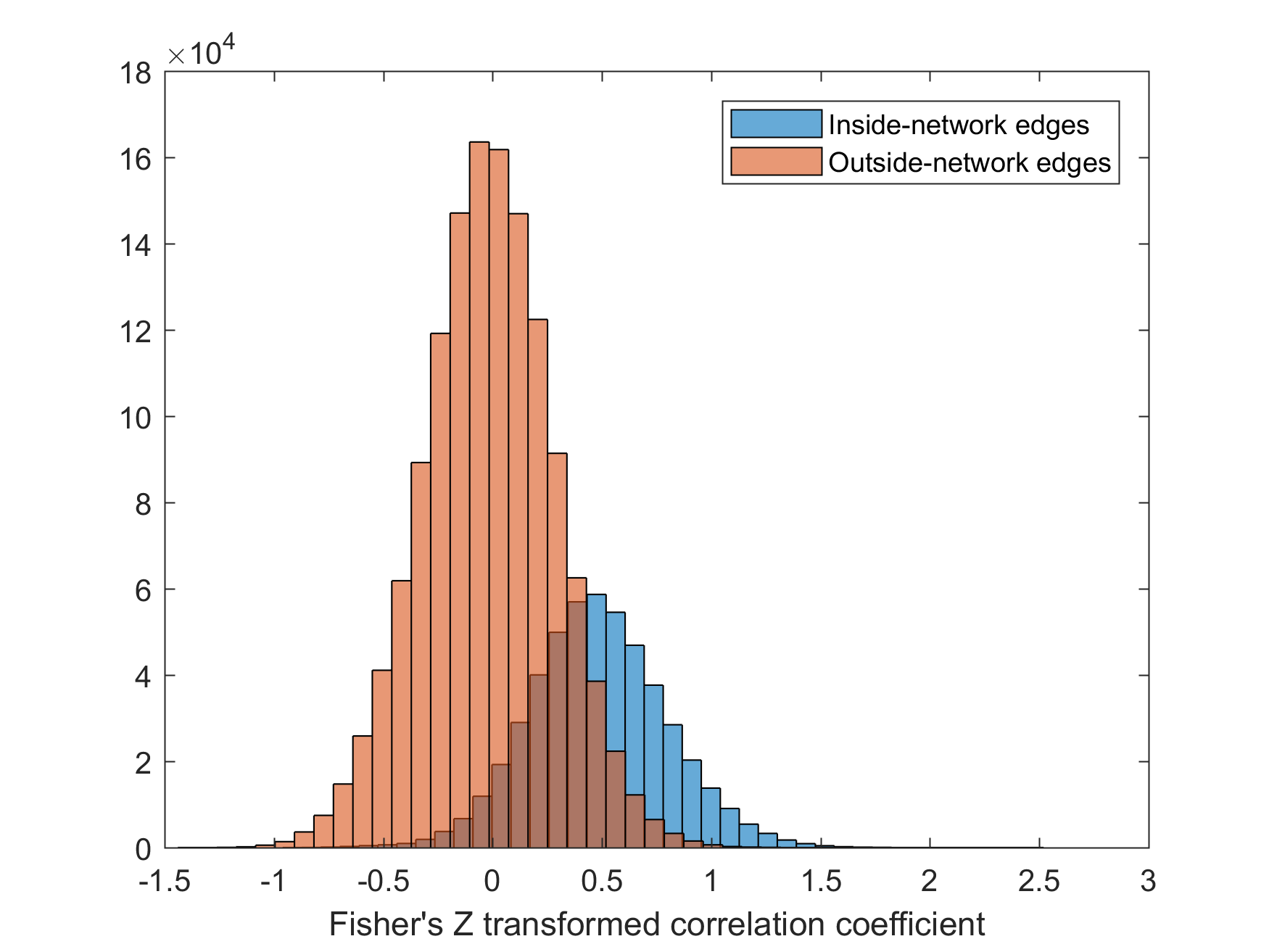}\label{fig3:f3}} 
  \hfill
  \subfloat[Estimated edge set $\hat{\textbf{E}}$]{\includegraphics[width=0.5\textwidth]{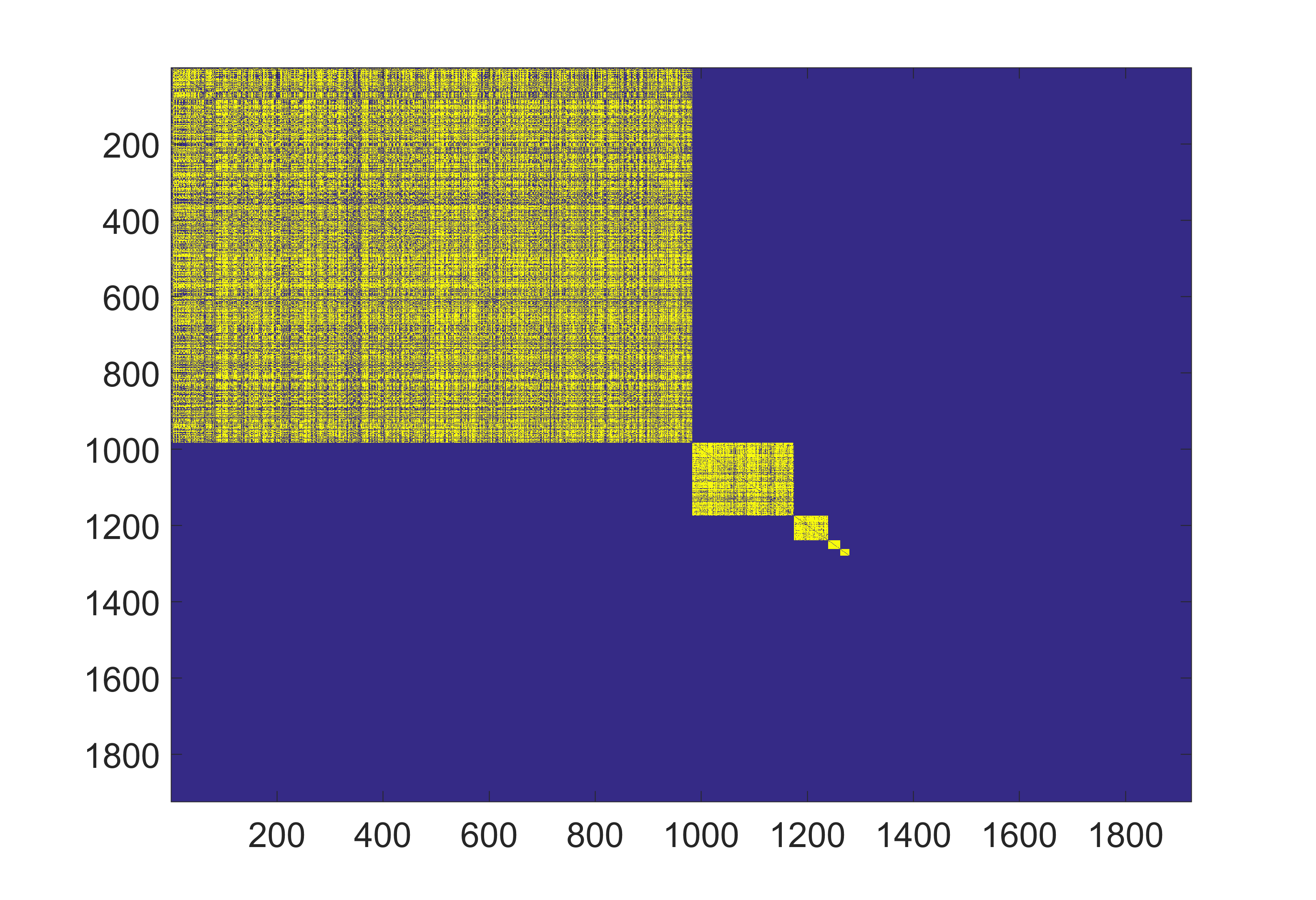}\label{fig3:f4}} 
  \caption{Application of the NICE to the example data set two. (a) is the sample correlation matrix; (b) demonstrates the latent  $G^1 \cup G^0$ mixture structure by reordering the variables in the heatmap; (c) shows the distributions of edges inside and outside the networks; (d) is the estimated $\hat{E}$ based on the NICE thresholding.}
\end{figure}

\subsection{Simulation Studies}
We conduct numerical studies to evaluate the performance of our approach, and compare it with several existing  methods. 
\subsubsection{Synthetic data sets}
We simulate each data set with $p=100$ variables, and thus $|V|=100$ and $|E|= {100 \choose 2}=4950 $. We assume that the correlation matrix includes two community networks, and the first include 15 nodes and the second 10 nodes. The induced networks are complete subgraphs (cliques) that all edges are connected within these two networks and no other edges are connected outside the two networks (Figure  \ref{fig1:f1}). Next, we permute the order of the nodes to mimic the practical data where the topological structure is latent. Figure  \ref{fig1:f2} represents the connected edges in the matrix. Let vector $\mathbf{x}_{p \times 1}^{k}$ follow a multivariate normal distribution, with zero mean and covariance matrix $ \boldsymbol \Sigma_{p \times p} $, and the sample size is $ n $. $\sigma_{i,j}$ is an entry at the $i$th row and $j$th column of $ \boldsymbol\Sigma $, $\sigma_{i,j}=1$ if $i=j$ (then $ \boldsymbol\Sigma = \textbf{R}$), and $\sigma_{i,j}=\rho$  if $e_{i,j} \in G_c$ (inside network edges) and $\sigma_{i,j}=0$ when $e_{i,j}\notin G_c$ (outside network edges). We simulate 100 data sets  at three different settings with different levels of signal to noise ratio (SNR) by using various sample sizes $n$ and  values of $\rho$. 
A larger sample size reduces the asymptotic variance of $\widehat \sigma_{i,j}$ and thus the noise level is lower; and a higher absolute value of $\rho$ represents higher signal level. A higher SNR leads to more distinct empirical distributions of $\widehat \sigma_{i,j}$ between inside network edges $e_{i,j}\in G_c$ and outside network edges $e_{i,j} \notin G_c$. Figure \ref{fig1:f3} demonstrates a calculated correlation matrix based on a simulated  data set. 

In our simulated data sets, 150 edges are connected and 4800 edges are unconnected, which together represent the graph edge skeleton $E$. For both precision matrix shrinkage and covariance matrix thresholding methods we treat a  non-zero  entry   $\widehat \delta_{i,j}=1$ (\citealp{MH}) as a connected edge. We summarize the false positive (FP) edges $\widehat \delta_{i,j}=1$ when the edge is not connected and ${e}_{i,j}\notin G_c$ and false negative (FN) edges $\widehat \delta_{i,j}=0$ when the edge is connected and ${e}_{i,j} \in G_c$. We compare FN and FP counts of each method by contrasting the estimated $\widehat E$ with the truth $E$. We compare our method with universal thresholding (Thresh), $glasso$, $el_1$ minimization for inverse matrix estimation (CLIME), and adaptive thresholding (AThres) by comparing the FP and FN  edges of estimating  the graph edge skeleton $E$ (\citealp{Bickel08}; \citealp{Friedman08}; \citealp{Cai11a}; \citealp{Cai11b}).

\begin{table}
	\caption{Median along with 25\% and 75\% quantiles of FP and FN}
	\label{quantiles}
	\begin{adjustbox}{width=1\textwidth}
		\small
		\begin{tabular}{r|c|rlrl|rlrl|rlrl}
\toprule
& &\multicolumn{4}{c|}{$\sigma=0.5$, $n=25$}&\multicolumn{4}{c|}{$\sigma=0.5$, $n=50$} & \multicolumn{4}	{c}{$\sigma=0.7$, $n=25$}\\
	Method			  &Tuning& \multicolumn{2}{c}{FP} & \multicolumn{2}{c|}{FN} & \multicolumn{2}{c}{FP} & \multicolumn{2}{c|}{FN} &\multicolumn{2}{c}{FP} & \multicolumn{2}{c}{FN}\\
			  			  &Par. &Med. & Quantiles &Med. & Quantiles &Med. & Quantiles &Med. & Quantiles &Med. & Quantiles &Med. & Quantiles \\
			  \midrule
			glasso&${0.1}$ &  1673&(1648,\  1702) & 59&(55,\  64)& 1621&(1591.5,\  1640) & 44&(40,\  46)& 1581.5&(1557,\  1606) & 45.5&(42,\  48) \\
			&${0.2}$ &  1008.5&(989,\  1025) & 59&(53.5,\  64.5)& 630&(610,\  644) & 38&(33.5,\  43) & 932.5&(920,\  955.5) & 36&(32,\  40) \\
			&${0.3}$ &  546&(529.5,\  560)&  56&(48,\  63.5)&151&(141,\  162.5) & 38&(30.5,\  43) & 500.5&(490,\  516) & 28&(23.5,\  33) \\
			&${0.4}$ &  211.5&(200.5,\  222.5) &  60&(50.5,\  72) &19&(16,\  21) & 48.5&(38,\  58) & 194&(186,\  204.5) &24.5&(20,\  29) \\
			&${0.5}$ &  51&(46,\  59)& 80.5&(66,\  96)& 1&(0,\  2) & 82.5&(67,\  96.5) &47&(41.5,\  54) & 28&(22.5,\  35) \\
			&${0.6}$ &  7&(5,\  10) &112.5&(97,\  125.5)& 0&(0,\  0) & 130&(118.5,\  137) & 6&(5,\  8.5) &41&(31,\  51) \\
			&${0.7}$ &  0&(0,\  1) &140&(131.5,\  146) & 0&(0,\  0) & 149&(147,\  150) &0&(0,\  1) & 75&(61.5,\  89)\\
			&${0.8}$ &  0&(0,\  0)& 149&(148,\  150)& 0&(0,\  0) & 150&(150,\  150) & 0&(0,\  0) &  127&(119.5,\  135)\\
			&${0.9}$ &  0&(0,\  0)& 150&(150,\  150)&0&(0,\  0) & 150&(150,\  150) & 0&(0,\  0) & 149&(149,\  150) \\
			&${1.0}$ 	 &  0&(0,\  0)& 150&(150,\  150)&0&(0,\  0) & 150&(150,\  150) & 0&(0,\  0) & 150&(150,\  150)\\[2mm]
			CLIME&${0.1}$ &  1082.5&(1047.5,\  1108) & 56&(48,\  64.5)& 993.5&(981,\  1024) & 39&(32,\  45.5)& 1054&(1021,\  1079) & 48.5&(40,\  56) \\
			&${0.2}$ &  353&(339.5,\  367.5) & 79.5&(69,\  87.5)& 241.5&(231.5,\  251.5)& 61&(54,\  67.5) & 345&(328,\  359) & 70&(59,\  78) \\
			&${0.3}$ &  63&(57,\  69)&  110&(98.5,\  115)&25&(22,\  29) & 92&(84.5,\  100) & 64&(59,\  68) & 98&(87,\  103) \\
			&${0.4}$ &  0&(0,\  1) &  140&(135,\  144) &0&(0,\  0) & 130&(124,\  135) & 0&(0,\  1) &134&(129,\  139) \\
			&${0.5}$ &  0&(0,\  0)& 150&(150,\  150)& 0&(0,\  0) & 150&(150,\  150) &0&(0,\  0) & 150&(150,\  150) \\[2mm]
		Thres&${0.1}$ &  2017.5&(1963.5,\  2067.5) & 0&(0,\  2)& 1978.5&(1944.5,\  2021.5) & 0&(0,\  0)& 2021.5&(1968.5,\  2061) & 0&(0,\  1) \\
			&${0.3}$ &  1292.50&(1252,\  1331)&  2&(0,\  5)&1249.5&(1220.5,\  1288.5) & 0&(0,\  0) & 1293.5&(1251,\  1341.5) & 1&(0,\  3) \\
			&${0.5}$ &  721.5&(699,\  752)& 5&(1,\  12)& 689&(673.5,\  721) & 0&(0,\  1) &722&(693,\  756) & 3&(1,\  10.5) \\
			&${0.7}$ &  344.5&(325,\  360) &14&(7,\  26.5)& 328.5&(311.5,\  349.5) & 1&(0,\  2) &342.5&(324,\  363) &10&(3,\  21.5) \\
			&${0.9}$ &  132&(121,\  143.5) &30&(18,\  45) & 129.5&(121,\  142) & 3&(1,\  7) &133&(123.5,\  146)& 24&(12,\  39.5)\\
			&${1.1}$ &  41.5&(35,\  46)& 55.5&(40,\  78.5)& 40.5&(36.5,\  47.5) & 10&(4.5,\  17) &40&(35.5,\  46.5) &  49.5&(28,\  63)\\
			&${1.3}$ &  9&(6,\  10)& 92&(74,\  112)&10&(8,\  12) & 25&(13,\  37) & 9&(6,\  11) & 78&(54.5,\  89) \\
			&${1.5}$ 	 & 1&(0,\  2)& 126&(112.5,\  137)&2&(1,\  3) & 50.5&(32.5,\  68) & 1&(0,\  2) & 106&(92.5,\  114)\\
			&${1.7}$ 	 & 0&(0,\  0)& 145&(138.5,\  148)&0&(0,\  0) & 85.5&(67,\  102.5) & 0&(0,\  0) & 132.5&(120.5,\  138.5)\\
			&${1.9}$ 	 & 0&(0,\  0)& 150&(149,\  150)&0&(0,\  0) & 120.5&(105,\  130) & 0&(0,\  0) & 147&(144,\  149)\\[2mm]
	AThres	&${0.3}$ &  2593&(2566.5,\  2627.5) & 2&(0,\  5)&2538.5&(2509.5,\  2571) & 0&(0,\  0)& 2594&(2563,\  2619) & 1&(0,\  3) \\
			&${0.5}$ &  1460&(1421.5,\  1486)&  5&(1,\  12)&1412.5&(1379.5,\  1440) & 0&(0,\  1) & 1453&(1419.5,\  1491) & 3&(1,\  10.5)\\
			&${0.7}$ &  691.5&(667,\  717)& 14&(7,\  26.5)& 668.5&(646,\  697) & 1&(0,\  2) &695.5&(665.5,\  720) & 10&(3,\  21.5) \\
			&${0.9}$ &  271.5&(258,\  291.5) &30&(18,\  45)& 265&(252,\  283.5) & 3&(1,\  7) &270.5&(255.5,\  288) &24&(12,\  39.5) \\
			&${1.1}$ &  83&(75,\  95) &55.5&(40,\  78.5) & 85&(75.5,\  95.5) & 10&(4.5,\  17) &82&(74,\  89.5)& 49.5&(28,\  63)\\
			&${1.3}$ &  18&(15,\  21)& 92&(74,\  112)&22&(18.5,\  25.5) &25&(13,\  37) &18&(14.5,\  22) &  78&(54.5,\  89)\\
			&${1.5}$ & 2&(1,\  4)& 126&(112.5,\  137)&4&(3,\  6) & 50.5&(32.5,\  68) & 3&(1,\  3) & 106&(92.5,\  114) \\
			&${1.7}$ 	 & 0&(0,\  0)& 145&(138.5,\  148)&0&(0,\  1) & 85.5&(67,\  102.5) & 0&(0,\  0) &132.5&(120.5,\  138.5)\\
			&${1.9}$ 	 & 0&(0,\  1)& 150&(149,\  150)&0&(0,\  0) & 120.5&(105,\  130) & 0&(0,\  0) &147&(144,\  149)\\[2mm]
			NICE & None &  44&(15,\  98) & 3&(0,\  27)& 11&(1,\  30) & 0&(0,\  4)& 32.5&(13.5,\  71) & 14&(4,\  38.5)
			\\ \bottomrule
		\end{tabular}
	\end{adjustbox}
\end{table}

\subsubsection{Simulation Study Results}
The simulation results are summarized in Table~\ref{quantiles}. Rather than selecting a single tuning parameter $\lambda$ for glasso and other methods by cross-validation, we  explore all possible choices within a reasonable range and use the one with best performance for comparison. Cross the 100 simulation data sets, we summarize the 25\%, 50\%, and 75\% quantiles  of the number of FP and FN edges to assess the performance of each method. The results show that the NICE algorithm outperforms the competing methods even when optimal tuning parameters are used (after comparing with the truth) for these methods. One possible reason could be the NICE algorithm thresholds the correlation matrix based on the topological structure rather than the a universal shrinkage or thresholding strategy.  More importantly, our approach is the only method can automatically detect the underlying  $G^1 \cup G^0$ mixture    topological structure. When the graph topological structure does not exist, the performance of all methods are similar across all settings. The matrix norm loss is not compared, because the community networks are small in size and norm comparison are likely determined by the false positive edges outside network communities.  We note that the methods with sparsity assumption (e.g. \textit{glasso} and CLIME) may miss many connected edges (false negative discovery rates are higher) even when small tuning parameter is used (false positive rates are high). Therefore, when a latent topological structure exists the sparsity assumption may not be valid because a cluster of features within a network are all correlated with each other and many of them can be conditionally independent. 

In summary, the numerical results demonstrate that our new method not only provides more accurate estimation of the correlation matrix and the edge set $E$ than the competing method, but also automatically detects the community networks where highly correlated edges distribute in an organized fashion.  
\section{Discussion and Conclusion}

We develop a NICE algorithm to bridge the correlation matrix estimation and graph topological structure detection via a flexible empirical Bayesian framework.  Recognizing the latent network topological structure not only can reveal underlying biological pathways, but also can guide the decision making procedure of regularization.

The latent $G^1 \cup G^0$ mixture graph structure exist widely in high-throughput biomedical data, however, the conventional network detection and clustering algorithms may not detect it due to the impact of  false positive noises. For instance, a few false positive edges may lead to detecting a large networks with low proportion of highly connected edges. The proposed penalized network estimation objective function can identify the mixture structure because it is less affected by false positive noises. Interestingly, we find that the number of networks is related to the penalty term because a larger $C$ generate many singletons. The optimization of the objective function is interestingly linked to the spectral clustering algorithms and the computational speed is affordable. 


Next, the new Bayes factor based thresholding approach naturally incorporates detected network topological structure from step one as prior knowledge. The updated thresholding values are determined by each edge's `location' on the detected graph topological `map'. Therefore, edges borrow strength with each other with higher precision based on detected topological structure, which also provides a flexible pathway to account for the dependency between edges. With additional information from the detected topological structure and appropriate modeling strategy, our new  thresholding approach reduces false positive and false negative  rates simultaneously when topological structures exist. Clearly, the performance of graph topological structure detection influences the accuracy of correlation matrix thresholding because it determines the empirical distributions of $z_{i,j}^{in}$ and  $z_{i,j}^{out}$ and thus $\widehat \theta_{in} $ and $\widehat \theta_{out} $. 
Therefore, the two steps of the NICE algorithm are seamlessly connected as the parsimonious property of the network detection ensures the efficiency and accuracy of the following regularization step. Edges outside networks are subject to more stringent thresholds whereas edges inside networks are less likely to be thresholded. This decision rule is data-oriented and determined by the latent spatial distributions of edges in the sample correlation matrix.  Last but not least, we develop theoretical results to prove the consistency of edge selection and estimation.



In our application, only positive (correlation) edges are distributed in an organized graph topology and the negative (correlation) edges are randomly distributed. Based on the network based thresholding,  negative (correlation) edges are thresholded. Our methods are ready to be extended to the scenario that negatively correlated edges show a organized topological structure. The numerical studies and example data  application have demonstrated excellent performance of the NICE algorithm regarding false positive/negative findings and latent network detection. The computational cost of NICE algorithm is low (for our simulation example the algorithm only takes 40 seconds using i7 CPU and 24G memory), and thus it is ready to scale up for larger data sets. In addition, the NICE algorithm is not restricted for multivariate Gaussian distributed data and it is straightforward to extend the sample correlation matrix to other sample metrics, for example maximal information coefficients (\citealp{Kinney14}) for continuous data and  polychoric correlation coefficient for categorical  data (\citealp{Bonett05}) because  graph topology oriented thresholding are based on the empirical distribution of the coefficients.

\section*{Acknowledgements}
The research is based upon work supported by the Office of the Director of National Intelligence (ODNI), Intelligence Advanced Research Projects Activity (IARPA), via  DJF-15-1200-K-0001725.
\appendix
\section{Appendix}

\subsection{A new algorithm to optimize the objective function \ref{fm1}}

The objective function 
$$ \underset{C,  \{G_c\} }{\argmax} \sum_{C=1}^C \exp\{ \log(\sum (w_{i,j}|e_{i,j} \in G_c))-\lambda_0 \log(|E_c|) \}$$ 
is non-convex and NP hard. We solve it in two steps.

$\lambda_0$ is between 0 and 1, by default we set $\lambda_0= 1/2$ because we aim to include most informative edges in .   Firstly, we optimize $\{G_c\} $ with given $C$:
\begin{align}
\label{rcut}
\begin{split}
 &\underset{\{G_c\} }{\argmax}\sum_{C=1}^C \exp\{ \log(\sum (w_{i,j}|e_{i,j} \in G_c))-\lambda_0 \log(|E_c|) \} \\
=&\underset{\{G_c\} }{\argmax} \sum_{C=1}^C \left(  \frac{\sum (w_{i,j}|e_{i,j} \in G_c)}{|E_c|}\right) ^{\lambda_0} \left( \sum (w_{i,j}|e_{i,j} \in G_c)\right)^{1-\lambda_0} \\
\doteq&\underset{\{G_c\} }{\argmax} \sum_{C=1}^C  \rho_{CC}|V_c|, \mbox{ when } \lambda_0=1/2, \rho_{CC}=  \sum (w_{i,j}|e_{i,j} \in G_c)/|E_c|\\
=&\underset{\{G_c\} }{\argmax} \sum (w_{i,j}|e_{i,j} \in G) /|V|  - \sum_{C=1}^C\sum_{C' \neq C} \rho_{CC'}(|V_c|+|V_{C'}|)\\
\Leftrightarrow &\underset{\{G_c\} }{\argmin} \sum_{C=1}^C\sum_{C' \neq C} \rho_{CC'}(|V_c|+|V_{C'}|)\\
= &\underset{\{G_c\} }{\argmin} \sum_{C=1}^C\sum_{C' \neq C} \frac{  \sum (w_{i,j}|i \in  G_c, j \in  G_{C'})}{|V_c||V_{C'}|}(|V_c|+|V_{C'}|) \\
= &\underset{\{G_c\} }{\argmin} \sum_{C=1}^C \frac{  \sum (w_{i,j}|i \in  G_c, j \notin  G_{C})}{|V_c| }
\end{split}
\end{align}
 
We solve objective function \ref{rcut} by using spectral clustering algorithm RatioCut (\citealp{Chen15}). 

Next, we select $C$ by grid searching that maximizes the criteria:

$$\sum_{C=1}^C \left(  \frac{\sum (w_{i,j}|e_{i,j} \in G_c)}{|E_c|}\right) ^{\lambda_0} \left( \sum (w_{i,j}|e_{i,j} \in G_c)\right)^{1-\lambda_0} $$
%
 At this step, a larger $\lambda_0$ often leads to detected subnetworks with higher proportion of more informative edges and smaller sizes whereas a smaller $\lambda_0$ often produces larger networCs including more informative edges in $G$. The iterations of the above two steps implement the optimization of \ref{fm1}.

When a more complex subgraph topological structure of  $G_c$ (e.g. bipartite subgraph) exists in $G$ instead of the default clique structure, advanced graph topology detection tools are needed (e.g. \citealp{Chen16}). The detected organized subnetworks (with more complex graph topological structures) can increase the objective function \ref{fm1} as the quality term increases and quantity term is almost unchanged. Therefore, the refined graph topological structure detection algorithms, for instance, K-partite, rich-club, and overlapped subgraphs could further assist to optimize the objective function. In future, more graph topological structure automatic detection tools will be developed, which will be compatible with the objective function \ref{fm1}.
\subsection{\textbf{W} matrix calculation}
Let $z_{i,j}$ be  the Fisher's Z transformed correlation coefficient $\widehat R_{i,j}$, for instance (Kendall's Tau or other pairwise relationship metrics could also be applied). We could simply let $w_{i,j}=z_{i,j}$ or further transform it to the probability scale. Assume that sample correlation coefficients  for all edges follow a mixture distribution $z_{i,j} \sim \pi_0 f_0(z_{i,j})+\pi_1 f_1(z_{i,j})$ where  $\pi_0+\pi_1=1$ (\citealp{Efron04}; \citealp{Wu06}; \citealp{Efron07}).   $f_1$ represents the distribution of correlations corresponding to the component of connected edges  $z_{i,j}|(\delta_{i,j}=0) \sim f_0(z_{i,j})$ , and $f_0$ for the unconnected edges $z_{i,j}|(\delta_{i,j}=1) \sim f_1(z_{i,j})$.   We adopt the empirical Bayes method to obtain {$\hat{\pi}_0, \hat{\pi}_1, \hat{f}_0, \hat{f}_1$},  ( \citealp{Efron07}) and then $w_{i,j}$ is the posterior probability that $z_{i,j}$ from the non-null component.  
\subsection{Convenient thresholding value calculation} 

We calculate the thresholding values for edges inside-networks or outside-networks separately, and these cut-offs can be linked to the overall local $fdr$ value. Therefore, the computation is more straightforward by using the following cut-offs.

An edge inside networks $z^{in}$  is truly connected if $fdr^{in}(z^{in}) \leq 1/(T+1)$,  where T is the threshold. Equivalently if $\frac{f_1(z)}{f_0(z)} \geq T \frac{\pi_0^{in}}{\pi_1^{in}}$, we consider the edge is connected  by using the fact below. 
\begin{align} \label{c1}
\begin{split} 
\frac{\pi_1^{in}f_1(z^{in})}{\pi_0^{in}f_0(z)} &= (1-fdr^{in}(z^{in}))/fdr^{in}(z^{in})\geq T \\
 \Rightarrow \frac{f_1(z^{in})}{f_0(z^{in})} & \geq T \frac{\pi_0^{in}}{\pi_1^{in}}
\end{split}   
\end{align}

The above cut-off can be linked to $fdr^{all}(z^{in}) $ by using the fact that
\begin{align} \label{c2}
\begin{split} 
&\frac{f_1(z^{in})}{f_0(z^{in})} \geq T \frac{\pi_0^{in}}{\pi_1^{in}} \\
  \Rightarrow  & \frac{f_1(z^{in})}{f_0(z^{in})} \frac{\pi_1^{all} }{\pi_0^{all} }  =(1-fdr^{all}(z^{in}))/fdr^{all}(z^{in}) \geq T \frac{\pi_0^{in}}{\pi_1^{in}}\frac{\pi_1^{all} }{\pi_0^{all} } \\
  \Rightarrow  & fdr^{all}(z^{in})=\frac{1}{T\frac{\pi_0^{in}}{\pi_1^{in}}\frac{\pi_1^{all} }{\pi_0^{all} } +1}
\end{split} 
\end{align}

For example, if $T=4, {\pi_0^{in}}/{\pi_1^{in}}=0.1$ and $ {\pi_1^{all} }/{\pi_0^{all} } =0.1$, then $fdr^{all}(z^{in})$=0.96 and most edges inside networks are considered as connected because the topological structure suggests that threshold is loose.

Similarly, for edges outside  networks $$fdr^{all}(z^{out})=\frac{1}{T\frac{\pi_0^{out}}{\pi_1^{out}}\frac{\pi_1^{all} }{\pi_0^{all} } +1}$$

if $T=4,  {\pi_0^{out}}/{\pi_1^{out}}=40$ and $ {\pi_1^{all} }/{\pi_0^{all} } =0.1$, then $fdr^{all}(out)$=0.06 and most edges outside the network are thresholded by using a more stringent cut-off. 

By using these overall $fdr$ based thresholds, the computation is faster. More importantly, we can note how topological structure and edge distributions can jointly impact the decision making process of the correlation matrix thresholding. When the data shows no network structure, for instance, ${\pi_0^{in}}/{\pi_1^{in}}= {\pi_0^{out}}/{\pi_1^{out}}=10$ and $ {\pi_1^{all} }/{\pi_0^{all} } =10$, then $fdr^{all}(in)=fdr^{all}(out)=0.2$. Our thresholding rule boils down to the universal thresholding rule.
\subsection{NICE algorithm}

The following is the detailed NICE algorithm.

\begin{algorithm}[ht!]
\caption{NICE algorithm}
\label{NICEalg}
\begin{algorithmic}[1]
\Procedure{NICE\textendash Algorithm}{}
\State Obtain the empirical Bayes fuzzy logic matrix $\textbf{W}=g(\widehat{ \textbf{R}})$;
\State Calculate the Laplacian matrix \textbf{L}=\textbf{D}-\textbf{W}
\For{cluster number  $C$ = $2:|V|-1$ }
\State Compute the first $C$ eigenvectors $[u_2, \cdots, u_{C}]$ of $L$, with eigenvalues ranked from the smallest;
\State Let $U={[u_2^T, \cdots, u_{C}^T]}$ be a $|V|\times C$ matrix containing all $C-1$ eigenvectors;
\State Perform K-means clustering algorithm on $U$ with number of clusters of $C$ to cluster $|V|$ nodes into $C$ networks;
\State Calculate the `quality and quantity' criterion for each $C$.
\EndFor
\State Adopt the clustering results using the $C$ of the maximum score of the `quality and quantity' criterion.
\State Identify the networks with significantly high proportion of correlated edges by using permutation test: for each detected community network $G_c$ in $G$

i) calculate the $T^0_c= -\log (1-\frac{1}{\Gamma(|E_c|)} \gamma( |E_c|,  \sum_{i,j \in G_c} -\log (W_{i,j})) $, where $\Gamma$ is the upper incomplete gamma function and $\gamma$ is the lower incomplete gamma function;

 ii) list all $\{W_{i,j}\}$ in \textbf{W} as a vector and shuffle the order of the vector and assemble the shuffled vector as a permuted $\textbf{W}^m$ for M (e.g. 10,000) times;
 
iii) calculate the maximum statistic $T^m_{max}$ for all detected communities in each iteration;

iv) calculate the percentile of $T^0_c$ in $\{T^m_{max}\}$, if it is less than the $\alpha$ level then the network $G_c$ is considered as a true community network.
\State Implement the topological structure oriented thresholding strategies for covariance entries inside and outside networks (see details in 2.2)
\EndProcedure
\end{algorithmic}
\end{algorithm}
 
\subsection{Proof of Lemma \ref{lem:prob_bound}}
\begin{proof}
For any $i, j$, let $Y_{c} = X_{i,k} X_{j,k}$. Then $Y_1,\ldots, Y_n$ are independent and identically distributed. By Condition \ref{con:data}, $\mathrm{P}[|Y_c| < M^2] = 1$. Define
$$z_{i,j} = g(Y_1, \ldots, Y_l, \ldots Y_n) =  \frac{1}{2}\log\left(\frac{1+\sum_{c\neq l}^n Y_c/n + Y_l/n}{1-\sum_{c\neq l}^n Y_c/n-Y_l/n}\right).$$
Then for all $l = 1,\ldots, n$, by Taylor expansion,we have
\begin{eqnarray*}
&&g(Y_1,\ldots, Y_l, \ldots, Y_n) - g(Y_1,\ldots, Y'_l, \ldots, Y_n) \\
&&= \sum_{c=1}^\infty \frac{\partial^C g}{\partial Y_l^C}(Y_1,\ldots, Y'_l,\ldots, Y_n)\frac{(Y_l - Y'_l)^C}{k!}
\end{eqnarray*}
where
$$
 \frac{\partial^C g}{\partial Y_l^C}(Y_1,\ldots, Y'_l,\ldots, Y_n) = \frac{(k-1)!}{2 n^C}\left( \frac{(-1)^{k+1}}{(1 + \widehat R_{i,j})^C}+ \frac{1}{(1 - \widehat R_{i,j})^C}\right),
$$
By Condition \ref{con:data}, $|R_{i,j}| < 1$ and strong law of large number, $\widehat R_{i,j} < 1$ with probability one. Thus, there exists $N > 0$ and $K_0>0$, for all $n>N$, we have
$$
\sup_{Y_1,\ldots, Y_n, Y_l'}|g(Y_1,\ldots, Y_l, \ldots, Y_n) - g(Y_1,\ldots, Y'_l, \ldots, Y_n)| \leq \frac{K_0}{n}
$$
By the McDiarmid inequality, for all $n \geq N$, we have 
$$\mathrm{P}[|z_{i,j}-\mE[z_{i,j}]| > \epsilon] \leq \exp\left(- \frac{2n\epsilon^2}{K^2_0}\right). $$
Thus, 
$$\mathrm{P}[\sqrt{n}|z_{i,j}-\mE[z_{i,j}]| > \epsilon] \leq \exp\left(- \frac{2}{K^2_0}\epsilon^2\right). $$
Taking $K = 2/K^2_0>0$ completes the proof. 
\end{proof}
\subsection{Proof of Lemma \ref{lem:norm_bound}}
\begin{proof}
When $e_{i,j} \in G$ and $\mu_{i,j}>0$, then 
\begin{eqnarray*}
\lefteqn{\mathrm{P}\left[\frac{\sum_{i',j'}\delta^0_{i',j'}\phi\{\sqrt{n}(z_{i,j}-\mu_{i',j'})\}}{\{p_n(p_n-1)/2- q_n\}\phi(\sqrt{n} z_{i,j})}\leq T\right] }\\
&\leq&\mathrm{P}\left[\frac{\phi\{\sqrt{n}(z_{i,j}-\mu_{i,j})\}}{\{p_n(p_n-1)/2- q_n\}\phi(\sqrt{n} z_{i,j})}\leq T\right] \\
& =&\mathrm{P}\left[ -(z_{i,j} - \mu_{i,j})^2 + z_{i,j}^{2(n)} \leq \frac{2}{n}[\log(T)+\log\{p_n(p_n-1)/2- q_n\}]\right]\\
&=&\mathrm{P}\left[  z_{i,j}   \leq \frac{1}{2}\mu_{i,j} + \frac{1}{\mu_{i,j} n}[\log(T)+\log\{p_n(p_n-1)/2- q_n\}]\right ]\\
&=&\mathrm{P}\left[ \sqrt{n}(z_{i,j}-\mu_{i,j})    \leq -\sqrt{n}\mu_{i,j}/2+ \frac{1}{\mu_{i,j}\sqrt{n}}\{\log(T)+\log\{p_n(p_n-1)/2- q_n\}\}\right]\\
&\leq &\mathrm{P}\left[  \sqrt{n} (z_{i,j}-\mu_{i,j})   \leq -\sqrt{n}\mu_{\inf}/2 +\frac{1}{\mu_{\sup}\sqrt{n}}\{\log(T)+\log(\{p_n(p_n-1)/2- q_n\}\}\right],
\end{eqnarray*}
where $\mu_{\sup} = \sup_{i,j}\{|\mu_{i,j}| : e_{i,j} \in G\}$ and $\mu_{\inf}$ is defined in Condition \ref{con:mu}.   Note that $\mE[z_{i,j}] = \mu_{i,j} + o(n^{-1/2})$. There exists $N_2$, for all $n>N_2$, such that 
$\sqrt{n}(\mu_{i,j} - \mE[z_{i,j}] ) < 1$ 
Then $- \sqrt{n}\mE[z_{i,j}] -1 < -\sqrt{n}\mu_{i,j} $ and thus, 
\begin{eqnarray*}
&&\mathrm{P}\left[  \sqrt{n} (z_{i,j}-\mu_{i,j})   \leq -\sqrt{n}\mu_{\inf}/2 +\frac{1}{\mu_{\sup}\sqrt{n}}\{\log(T)+\log(p_n(p_n-1)/2-q_n)\}\right]\\
&&\leq \mathrm{P}\left[\sqrt{n}  (z_{i,j}-\mE[z_{i,j}]) \leq -\sqrt{n}\mu_{\inf}/2 +\frac{1}{\mu_{\sup}\sqrt{n}}\{\log(T)+\log(p_n(p_n-1)/2-q_n)\}+1\right] 
\end{eqnarray*}
By Condition \ref{con:mu}, we have $\mu_{\inf} = c_0 n^{-1/2+\tau}$ with $\tau>0$, then $\mu_{\sup} > c_0 n^{-1/2+\tau} $.  Also, by Conditions \ref{con:p_q} and \ref{con:pi}, we have $\log\{p_n(p_n-1)/2-q_n\}= o(n^{2\tau})$, then there exists $N_1$ such that for all $n > N_1$ we have $n^{-\tau}\log\{T (p_n(p_n - 1)/2 - q_n)\} < c^2_0 n^{\tau}/8$ and $n^{\tau}>8/c_0$. By Lemma \ref{lem:prob_bound}, 
\begin{eqnarray*}
\lefteqn{\mathrm{P}\left[\frac{\sum_{i',j'}\delta^0_{i',j'}\phi\{\sqrt{n}(z_{i,j}-\mu_{i',j'})\}}{\{p_n(p_n - 1)/2 - q_n\}\phi(\sqrt{n} z_{i,j})}\leq T\right] }\\
&\leq & \mathrm{P}\left[ \sqrt{n} (z_{i,j}-\mu_{i,j}) \leq - \frac{c_0}{4} n^{\tau}\right] \leq \exp\left(-\frac{c_0^2 K n^{2\tau}}{16}\right).
\end{eqnarray*}
When $\mu_{i,j}<0$, based on  similar arguments, we have 
\begin{eqnarray*}
\lefteqn{\mathrm{P}\left[\frac{\sum_{i',j'}\delta^0_{i',j'}\phi\{\sqrt{n}(z_{i,j}-\mu_{i',j'})\}}{\{p_n(p_n - 1)/2 - q_n)\phi(\sqrt{n} z_{i,j}\}}\leq T\right] }\\
&\leq & \mathrm{P}\left[ -\sqrt{n} (z_{i,j}-\mu_{i,j}) \leq - \frac{c_0}{4} n^{\tau}\right] \leq \exp\left(-\frac{c_0^2 K n^{2\tau}}{16}\right).
\end{eqnarray*}
Taking $C_1 = c_0^2 K /16$ completes the proof for the case when $e_{i,j} \in G$.

When $e_{i,j} \notin G$, then 
\begin{eqnarray*}
\lefteqn{\mathrm{P}\left[\frac{\sum_{i',j'}\delta^0_{i',j'}\phi\{\sqrt{n}(z_{i,j}-\mu_{i',j'})\}}{(p_n(p_n - 1)/2 - q_n)\phi(\sqrt{n}z_{i,j})}> T\right] }\\
&\leq& \mathrm{P}\left[\frac{q_n\phi\{\sqrt{n}(z_{i,j}-\mu_{\inf}/3)\}}{(p_n(p_n - 1)/2 - q_n)\phi(\sqrt{n} z_{i,j})}> T, |z_{i,j}|\leq \mu_{\inf}/3\right] \\
&&\qquad \qquad + \mathrm{P}\left[\frac{\sum_{i',j'}\delta^0_{i',j'}\phi\{\sqrt{n}(z_{i,j}-\mu_{i',j'})\}}{(p_n(p_n - 1)/2 - q_n)\phi(\sqrt{n}z_{i,j})}> T, |z_{i,j}|> \mu_{\inf}/3\right]\\
&\leq&\mathrm{P}\left[ \sqrt{n}z_{i,j}  > \sqrt{n}(\mu_{\inf}/6)+ \frac{3}{\mu_{\inf}\sqrt{n}}\log\left(T\times \frac{p_n(p_n - 1)/2-q_n}{q_n}\right)-1\right] + \mathrm{P}\left[|z_{i,j}|> \mu_{\inf}/3\right]. 
\end{eqnarray*}
By Condition \ref{con:pi}, $\lim_{n\to\infty}\log\{(p_n(p_n - 1)/2 - q_n)/q_n\}=\pi_0/(1-\pi_0)$, there exists $N_0$ such that for all $n > N_0$, $\log\{T (p_n(p_n - 1)/2 - q_n)/q_n\} > 0$ and $n^{\tau} >  12/c_0$. Thus, 
\begin{eqnarray*}
\lefteqn{\mathrm{P}\left[\frac{\sum_{i',j'}\delta^0_{i',j'}\phi\{\sqrt{n}(z_{i,j}-\mu_{i',j'})\}}{(p_n(p_n - 1)/2 - q_n)\phi(\sqrt{n}z_{i,j})}> T\right] }\\
&\leq &\mathrm{P}\left[ \sqrt{n}z_{i,j}  > c_0n^{\tau}/12 \right] + \mathrm{P}\left[ \sqrt{n}|z_{i,j}|  > c_0 n^{\tau}/3 \right] \leq 3\exp\left(-\frac{c_0^2 K}{144}n^{2\tau}\right). 
\end{eqnarray*}
Taking $C_0 = c_0^2K/144$, $C_2 = 3$ and $N_T = \max\{N_0, N_1, N_2\}$ completes the proof for all the cases. 
\end{proof}

\subsection{Proof Lemma \ref{lem:e_bound}}
\begin{proof}
Since $T >(1-\pi_0)/\pi_0$. Then there exists $\epsilon > 0$ such that $T-\epsilon > (1-\pi_0)/\pi_0$. 
By Lemma \ref{lem:f_ratio} and Condition \ref{con:pi}, there exists $N_1 > 0$, for all $n > N_1$ such that 
$$\frac{\pi_1f_1(z_{i,j})}{\pi_0 f_0(z_{i,j})}  > \frac{\sum_{i',j'}\delta^0_{i',j'}\phi\{\sqrt{n}(z_{i,j}-\mu_{i',j'})\}}{(p_n(p_n - 1)/2 - q_n)\phi(\sqrt{n}z_{i,j})}  - \epsilon,$$
$$\frac{\pi_1f_1(z_{i,j})}{\pi_0 f_0(z_{i,j})}  < \frac{\sum_{i',j'}\delta^0_{i',j'}\phi\{\sqrt{n}(z_{i,j}-\mu_{i',j'})\}}{(p_n(p_n - 1)/2 - q_n)\phi(\sqrt{n}z_{i,j})}  + \epsilon.$$
Then by Lemma \ref{lem:norm_bound}, when $e_{i,j} \in G$, then $\delta_{i,j}^0 = 1$ and
\begin{eqnarray*}
\lefteqn{\mathrm{P}[\widetilde\delta_{i,j}(T) = 0]= \mathrm{P}\left[\frac{\pi_1f_1(z_{i,j})}{\pi_0 f_0(z_{i,j})} \leq T\right]}\\
&&\leq \mathrm{P}\left[\frac{\sum_{i',j'}\delta^0_{i',j'}\phi\{\sqrt{n}(z_{i,j}-\mu_{i',j'})\}}{(p_n(p_n - 1)/2 - q_n)\phi(\sqrt{n}z_{i,j})} \leq T+\epsilon\right] \leq \exp(-C_1 n^{2\tau}).
\end{eqnarray*}
when $e_{i,j} \notin G$, then $\delta_{i,j}^0 = 0$ and
\begin{eqnarray*}
\lefteqn{\mathrm{P}[\widetilde\delta_{i,j}(T) = 1]= \mathrm{P}\left[\frac{\pi_1f_1(z_{i,j})}{\pi_0 f_0(z_{i,j})} > T\right]}\\
&&\leq \mathrm{P}\left[\frac{\sum_{i',j'}\delta^0_{i',j'}\phi\{\sqrt{n}(z_{i,j}-\mu_{i',j'})\}}{(p_n(p_n - 1)/2 - q_n)\phi(\sqrt{n}z_{i,j})} > T-\epsilon\right] \leq C_2 \exp(-C_0 n^{2\tau}).
\end{eqnarray*}
Taking $C_3 = \max\{1,C_2\}$ and $C_4 = \min\{C_0,C_1\}$, thus, 
$$\Pr[\widetilde\delta_{i,j}(T) \neq \delta^0_{i,j}] \leq C_3 \exp(-C_4 n^{2\tau}).$$
\end{proof}
\subsection*{Proof of Lemma \ref{lem:e_hat_bound}}
\begin{proof}
Since $T>(1-\pi_0)/\pi_0$, then there exists $\epsilon > 0$ such that $T -\epsilon > (1-\pi_0)/\pi_0$. 
By Condition \ref{con:est}, there exists $N_0>0$, for all $n > N_0$ and all $i, j$ such that
$${\widehat \pi_1\widehat f_1(z_{i,j}) \over \widehat \pi_0\widehat f_0(z_{i,j})} > { \pi_1 f_1(z_{i,j}) \over  \pi_0 f_0(z_{i,j})} - \epsilon, \quad \mbox{and} \quad {\widehat \pi_1\widehat f_1(z_{i,j}) \over \widehat \pi_0\widehat f_0(z_{i,j})} < { \pi_1 f_1(z_{i,j}) \over  \pi_0 f_0(z_{i,j})} + \epsilon.$$
By Lemma \ref{lem:e_hat_bound}, When $e_{i,j} \in G$, then $\mathrm{P}[\widehat\delta_{i,j}(T)=0] \leq \mathrm{P}[\widetilde\delta_{i,j}(T+\epsilon)=0] \leq C_3 \exp(-C_4 n^{2\tau})$, and when $e_{i,j} \notin G$, then  $\mathrm{P}[\widehat\delta_{i,j}(T)=1] \leq \mathrm{P}[\widetilde\delta_{i,j}(T-\epsilon)=1] \leq C_3 \exp(-C_4 n^{2\tau})$. 
\end{proof}

\subsection{Proof of Theorem \ref{thm:selection_consistency}}
\begin{proof}
By Lemma \ref{lem:e_hat_bound} and the Bonferroni inequality,
\begin{eqnarray*}
\lefteqn{\mathrm{P}\{\widehat\bfDelta(T) \neq \bfDelta_0\} =\mathrm{P}\left[\bigcup_{i<j} \{\widehat \delta_{i,j} (T)\neq \delta^0_{i,j}\} \right]}\\
&\leq& \sum_{1\leq i<j \leq p_n}  \mathrm{P}\left[\widetilde\delta_{i,j}(T) \neq \delta^0_{i,j} \right] \leq \frac{C_3}{2}p_n(p_n - 1)\exp\left(- C_4 n^{2\tau}\right). 
\end{eqnarray*}
By Condition \ref{con:p_q}, $\lim_{n\to\infty}p_n(p_n - 1)\exp\left(- C_4 n^{2\tau}\right) = 0$. This completes the proof. 
\end{proof}
\subsection{Proof of Theorem \ref{thm:estimation_consistency}}
\begin{proof}
Note that 
\begin{eqnarray*}
\lefteqn{\mathrm{P}[\|\mathrm{NICE}(\widehat\bfR; T) - \bfR\|_{\infty} > \epsilon]}\\
&&= \mathrm{P}[\|\mathrm{NICE}(\widehat\bfR; T) - \bfR\|_{\infty} > \epsilon; \widehat\bfDelta(T)  = \bfDelta_0] + \mathrm{P}[\|\mathrm{NICE}(\widehat\bfR; T) - \bfR\|_{\infty} > \epsilon; \widehat\bfDelta(T) \neq \bfDelta_0]\\
&&\leq \mathrm{P}\left[\max_{i<j, e_{i,j}\in G} |\widehat R_{i,j} - R_{i,j}|> \epsilon\right] + \mathrm{P}[\widehat\bfDelta(T) \neq \bfDelta_0] 
\end{eqnarray*}
By Bonferroni inequality and Hoeffding's inequality, we have 
\begin{eqnarray*}
\mathrm{P}\left[\max_{i<j: e_{i,j}\in G} |\widehat R_{i,j} - R_{i,j}|> \epsilon\right] \leq \sum_{i<j: e_{i,j}\in G} \mathrm{P}[|\widehat R_{i,j} - R_{i,j}|> \epsilon] \leq  q_n \exp\left(- \frac{n \epsilon^2}{2M^4}\right).
\end{eqnarray*}
By Conditions \ref{con:p_q} and \ref{con:pi}, $q_n = o(n^{2\tau})$. Since $0<\tau < 1/2$, then $\lim_{n\to\infty}q_n \exp\left(-n \epsilon^2/2M^4\right) = 0$. 
By Theorem \ref{thm:selection_consistency}, we have $\lim_{n\to\infty} \mathrm{P}[\widehat\bfDelta(T) \neq \bfDelta_0]  = 0$. This completes the proof. 

\end{proof}

\subsection{Proof of Theorem \ref{thm:FPN}}
\begin{proof}
Applying the universal decision rule with $z_0$ as threshold:  

\begin{align}
E(\#    FP)=m \int_{z_0}^ \infty \frac{\pi_0 f_0(z)}{f(z)} f(z) dz =m \pi_0 F_0(z_0)=m \omega \pi^{in}_0 F_0(z_0) + m(1-\omega)\pi^{out}_0 F_0(z_0)
\label{a1}
\end{align}
where $\int_{z_0}^ \infty f= F(z_0)$ and $m$ is the total number of edges $m=|E|$.

For edges in communities:
 \begin{align}
 E^{in}(\#    FP)=\omega m \int_{z_{in}}^ \infty \frac{\pi^{in}_0f^{in}_0(z)}{ f^{in}(z)} f^{in}(z) dz =\omega m \pi^{in}_0 F^{in}_0(z_{in}) 
\label{a2}
\end{align}

For edges outside communities:
\begin{align}
  E^{out}(\#    FP)=(1-\omega)m \int_{z_{0,out}}^ \infty \frac{\pi^{out}_0 f^{out}_0(z)}{ f^{out}(z)} f^{out}(z) dz =(1-\omega)m \pi^{out}_0 F^{out}_0(z_{0,out}) 
\label{a3}
\end{align}

where $z_{0,in}<z_0<z_{0,out} $, and $F^{out}_0(z) =F^{in}_0(z)=F _0(z)$.  There we expect $E(\#    FP)$ (\ref{a1})$>E^{in}(\#    FP)+E^{out}(\#    FP)$   (\ref{a2} $+$ \ref{a3}) if  

\begin{align} \label{a4}
\begin{split}
&  -\omega m \pi^{in}_0 (F_0(z_{in}) - F_0(z_0))  +(1-\omega)m \pi^{out}_0 (F_0(z_0) - F_0(z_{0,out}))>0,  \\
&  \Leftrightarrow  \frac{ F_0(z_0) - F_0(z_{0,out}) }{ F_0(z_{0,in}) - F_0(z_0) } > \frac{\omega  \pi^{in}_0}{(1-\omega)  \pi^{out}_0}
\end{split}  
\end{align}


We further calculate the expected number of true positive (TP) edges using universal threshold and in/out communities to evaluate the power of our network based thresholding.  

Applying the universal decision rule with $z_0$ as threshold:  

\begin{align}
E(\#    TP)=m \int_{z_0}^ \infty \frac{\pi_1 f_1(z)}{f(z)} f(z) dz =m \pi_1 F_0(z_0)=m \omega \pi^{in}_1 F_1(z_0) + m(1-\omega)\pi^{out}_1 F_1(z_0)
\label{b1}
\end{align}

For edges in communities:
 \begin{align}
 E^{in}(\#    \  TP)=\omega m \int_{z_{in}}^ \infty \frac{\pi^{in}_1f^{in}_1(z)}{ f^{in}(z)} f^{in}(z) dz =\omega m \pi^{in}_1 F^{in}_1(z_{in}) 
\label{b2}
\end{align}

For edges outside communities:
\begin{align}
  E^{out}(\#    \ TP)=(1-\omega)m \int_{z_{out}}^ \infty \frac{\pi^{out}_1f^{out}_1(z)}{ f^{out}(z)} f^{out}(z) dz =(1-\omega)m \pi^{out}_1 F^{out}_1(z_{out}) 
\label{b3}
\end{align}

where $z_{0,in}<z_1<z_{0,out} $, and $F^{out}_1(z) =F^{in}_1(z)=F _1(z)$.  There we expect $E(\#    TP)$ (\ref{b1})$<E^{in}(\#   TP)+E^{out}(\#   TP)$   (\ref{b2} $+$ \ref{b3}) (i.e.  $E(\#    FN)$  $>E^{in}(\#   FN)+E^{out}(\#   FN)$ ) if  

\begin{align} \label{b4}
\begin{split}
&  -\omega m \pi^{in}_0 (F_1(z_{in}) - F_1(z_0))  +(1-\omega)m \pi^{out}_1 (F_1(z_0) - F_1(z_{0,out}))<0,  \\
&  \Leftrightarrow  \frac{ F_1(z_0) - F_1(z_{0,out}) }{ F_1(z_{in}) - F_1(z_0) } < \frac{\omega  \pi^{in}_1}{(1-\omega)  \pi^{out}_1}
\end{split}  
\end{align}
\end{proof}
Condition   \ref{con:fp}   is generally true because our network detection algorithm (including the `quality and quantity criterions') and permutation test ensure the communities have large proportions of highly correlated edges.  We have run numerous empirical experiments and the results further verify this claim. If the assumption of network induced correlation matrix is true, the $C$ selection procedure chooses the parameter to optimize the objective function of step one that  reduce false positive findings and improve power simultaneously.

%
%
%
%
%

\end{document}

%% file: def.tex
\newcommand{\cm}[1]{\ignorespaces}

\def\bfR{\mathbf R}

\def\bfX{\mathbf X}

\def\bfM{\mathbf M}

\def\mVar{\mathrm{Var}}

\def\mE{\mathrm{E}}

\def\mF{\mathrm{F}}

\def\mbR{\mathbb R}

\def\md{\mathrm d}

\def\mE{\mathrm{E}}

\newtheorem{lemma}{\sc Lemma}[section]
\newtheorem{theorem}{\sc Theorem}

\newtheorem{condition}{Condition}[section]